\documentclass[aps,prapplied,twocolumn,nobibnotes,superscriptaddress,notitlepage, longbibliography]{revtex4-2}
\usepackage{graphicx}
\usepackage{upgreek}
\usepackage{amssymb,amsmath}
\usepackage{graphicx}
\usepackage{natbib}
\usepackage{mathrsfs}
\usepackage{overpic}
\usepackage{wrapfig}
\usepackage{braket}
\usepackage{siunitx}
\usepackage{gensymb}
\usepackage{booktabs}
\usepackage{xcolor}
\usepackage{color}
\usepackage[letterpaper,textwidth=7in,top=.75in,bottom=.75in]{geometry}
\definecolor{darkblue}{rgb}{0.0 0.0 0.78}
\definecolor{darkred}{rgb}{0.5 0.0 0.0}

\usepackage{ulem}
\newcommand{\RNC}[1]{\MakeUppercase{\romannumeral #1}}

\newcommand{\huPhys}{Department of Physics, Harvard University, Cambridge, Massachusetts 02138, USA}
\newcommand{\mitPhys}{Department of Physics, Massachusetts Institute of Technology, Cambridge, Massachusetts 02139, USA}
\newcommand{\cfa}{Harvard-Smithsonian Center	for	Astrophysics,	Cambridge,	Massachusetts	02138,	USA}
\newcommand{\cbs}{Center	for	Brain	Science,	Harvard	University,	Cambridge,	Massachusetts	02138,	USA}
\newcommand{\lincoln}{Lincoln Laboratory, Massachusetts Institute of Technology, Lexington, Massachusetts 02421, USA}
\newcommand{\ethzurich}{Department of Physics, ETH Zurich, 8093 Zurich, Switzerland}
\newcommand{\MaryPhys} {Department of Physics, University of Maryland, College Park, MD 20742, USA}
\newcommand{\MaryECE} {Department of Computer and Electrical Engineering, University of Maryland, College Park, MD 20742, USA}
\newcommand{\MaryQTC} {Quantum Technology Center, University of Maryland, College Park, MD 20742, USA}

\begin{document}

\title{NV-Diamond Magnetic Microscopy using a Double Quantum 4-Ramsey Protocol}
\date{\today}

\author{Connor A. Hart}  
\thanks{These authors contributed equally to this work}
\affiliation{\huPhys}
\affiliation{\MaryPhys}
\affiliation{\MaryECE}
\affiliation{\MaryQTC}

\author{Jennifer M. Schloss} 
\thanks{These authors contributed equally to this work}
\affiliation{\lincoln}
\affiliation{\mitPhys} 
\affiliation{\cbs}

\author{Matthew J. Turner} 
\affiliation{\huPhys}
\affiliation{\MaryPhys}
\affiliation{\MaryECE}
\affiliation{\MaryQTC}
\affiliation{\cbs}

\author{Patrick J. Scheidegger} 
\affiliation{\ethzurich}

\author{Erik Bauch} 
\affiliation{\cfa}

\author{Ronald L. Walsworth}
\thanks{walsworth@umd.edu}
\affiliation{\huPhys}
\affiliation{\MaryPhys}
\affiliation{\MaryECE}
\affiliation{\MaryQTC}
\affiliation{\cbs} 
\affiliation{\cfa}

\begin{abstract}
We introduce a double quantum (DQ) 4-Ramsey measurement protocol that enables wide-field magnetic imaging using nitrogen vacancy (NV) centers in diamond, with enhanced homogeneity of the magnetic sensitivity relative to conventional single quantum (SQ) techniques. The DQ 4-Ramsey protocol employs microwave-phase alternation across four consecutive Ramsey (4-Ramsey) measurements to isolate the desired DQ magnetic signal from any residual SQ signal induced by microwave pulse errors. In a demonstration experiment employing a 1-$\upmu$m-thick NV layer in a macroscopic diamond chip, the DQ 4-Ramsey protocol provides volume-normalized DC magnetic sensitivity of $\eta^\text{V}=34\,$nT\,Hz$^{-1/2}\,\upmu$m$^{3/2}$ across a \SI{125}{\micro\meter}$\times$\SI{125}{\micro\meter} field of view, with about 5$\times$ less spatial variation in sensitivity across the field of view compared to a SQ measurement. The improved robustness and magnetic sensitivity homogeneity of the DQ 4-Ramsey protocol enable imaging of dynamic, broadband magnetic sources such as integrated circuits and electrically-active cells.
\end{abstract}

\maketitle
\section{Introduction}

Nitrogen-vacancy (NV) color centers in diamond constitute a leading quantum sensing platform, with particularly diverse applications in magnetometry~\cite{Rondin2014}. The negatively-charged NV$^\text{-}$ center has an electronic spin-triplet ground state with magnetically-sensitive spin resonances, offers all-optical spin-state preparation and readout under ambient conditions, and can be engineered at suitably high densities in favorable geometries~\cite{Doherty2013,barry_sensitivity_2019}. These properties make ensembles of NV$^\text{-}$ centers particularly advantageous for wide-field magnetic microscopy of physical and biological systems with micrometer-scale spatial resolution, a modality known as the quantum diamond microscope (QDM)~\cite{Levine2019}. QDM applications to date include imaging magnetic fields from remnant magnetization in geological specimens~\cite{Glenn2017}, domains in magnetic memory~\cite{Simpson2016}, iron mineralization in chiton teeth~\cite{McCoey2020}, current flow in graphene devices~\cite{Tetienne2016,Ku2020} and integrated circuits~\cite{ICpaper}, populations of living magnetotactic bacteria~\cite{LeSage2013}, and cultures of immunomagnetically labeled tumor cells~\cite{Glenn2015}. 

Despite this progress, QDM magnetic imaging applications have been largely restricted to mapping of static magnetic fields exceeding several microtesla due to shortcomings of conventional single quantum (SQ) magnetometry. SQ schemes sense changes in the frequency or phase accumulation between two sublevels with difference
in spin projection quantum number $\Delta m = 1$. In particular, the sensitivity of QDMs using continuous-wave optically detected magnetic resonance (CW-ODMR) is impaired by competing effects of the optical and microwave (MW) control fields applied during the sensing interval~\cite{Levine2019,barry_sensitivity_2019}. Pulsed-ODMR schemes, which separate the optical spin-state preparation and readout from the MW control and sensing interval, offer improved sensitivity, but cannot exceed the performance achievable with SQ Ramsey magnetometry~\cite{barry_sensitivity_2019}.

Furthermore, any SQ magnetometry scheme is vulnerable to diamond crystal stress inhomogeneities and temperature variations, which shift and broaden the NV$^\text{-}$ spin resonances. Such stress gradients are particularly pernicious for QDM applications, with typical gradient magnitudes comparable to NV$^\text{-}$ resonance linewidths ($0.1\!-\!1\,$MHz) and spatial structure spanning the submicron to millimeter scales~\cite{StrainPaper}. Stress-induced resonance shifts or broadening may be mistaken for magnetic signals of interest. Stress gradients can also degrade per-pixel sensitivity and sensitivity homogeneity across an image. While protocols such as sequentially sampling the ODMR spectrum at multiple frequencies~\cite{Glenn2017} or employing four-tone MW control~\cite{kazi2020, Fescenko2019} can separate magnetic and non-magnetic signals, the worsened and inhomogeneous magnetic sensitivity caused by stress gradients remains unaddressed.

\begin{figure}[ht]
  \centering
  \includegraphics[width=8.5cm]{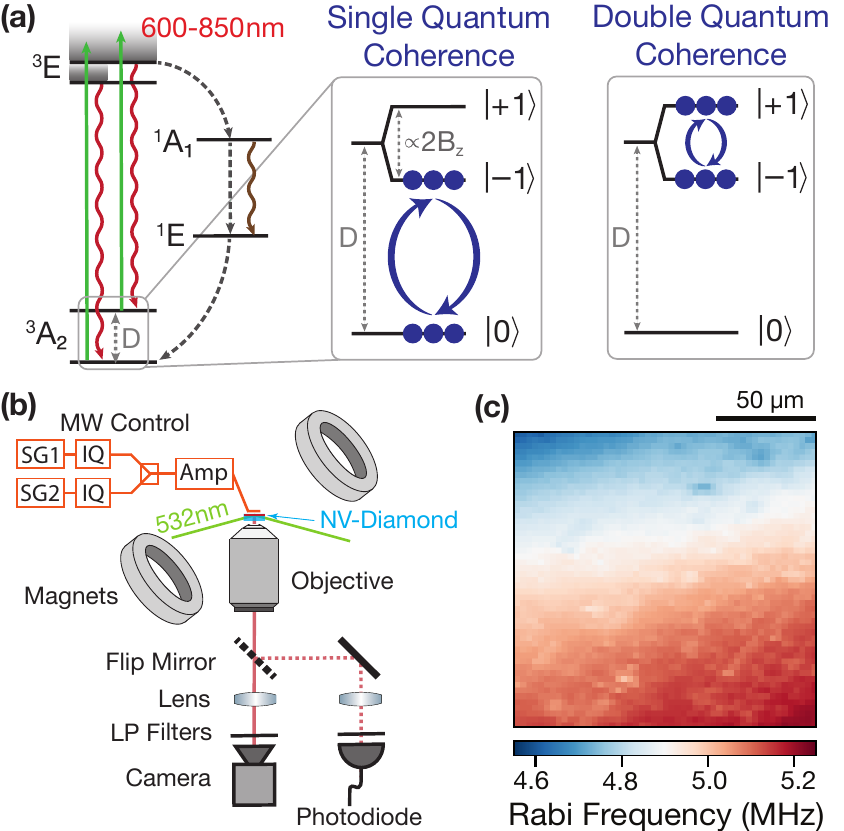} 
  \caption{
  \textbf{The NV$^\text{-}$ Energy Level Diagram and Experimental Apparatus. (a)} Energy level diagram for the negatively-charged nitrogen vacancy (NV$^\text{-}$) in diamond with zero field splitting $D$ between the ground state spin levels $|m_s\!=\!0\rangle$ and $|m_s=\!\pm\!1\rangle$. The enlarged views depict single quantum (SQ) and double quantum (DQ) coherence. To induce a DQ coherence, the $|0\rangle\! \rightarrow\!|\pm1\rangle$ spin transitions are simultaneously irradiated with a two-tone resonant microwave (MW) pulse. 
  \textbf{(b)} An overview of the QDM apparatus including \SI{532}{nm} excitation of a micron-scale layer of NV centers in a macroscopic diamond chip, using total internal reflection (TIR). NV fluorescence is collected using a 20$\times$ objective onto a camera or photodiode. 647$\,$nm and 532$\,$nm long-pass (LP) optical filters partially isolate NV$^\text{-}$ fluorescence from background NV$^\text{0}$ fluorescence. MW control fields are synthesized using two signal generators with phase control on both tones and applied via a millimeter-scale shorted coaxial loop. A bias magnetic field of \SI{5}{\milli\tesla} is aligned with NV centers oriented along a single crystallographic axis. 
  \textbf{(c)} Typical NV$^\text{-}$ Rabi frequency variation across a \SI{125}{\micro\meter}$\times$\SI{125}{\micro\meter} field of view. The effects of inhomogeneous, stress-induced NV$^\text{-}$ resonance shifts on the Rabi frequency are visible in addition to a quasi-linear Rabi gradient  due to spatial variation in the MW amplitude. 
  }
\label{fig:fig1}
\end{figure}

Here, we demonstrate a double quantum (DQ) 4-Ramsey protocol that overcomes the shortcomings of SQ CW- and pulsed-ODMR measurement techniques. This protocol expands upon the advantageous pulsed Ramsey scheme, which temporally separates the spin state control, optical readout, and sensing intervals. The scheme thus enables use of  increased laser and MW intensity compared to CW-ODMR, allowing for improved measurement contrast and higher fluorescence count rates without broadening the NV$^\text{-}$ spin resonances. Furthermore, the protocol exploits the benefits of DQ coherence magnetometry, which leverages a double quantum superposition of the $m_s=|\pm1\rangle$ ground-state sublevels, to cancel common-mode resonance shifts and broadening from stress, electric fields, and temperature variations~\cite{P1DQ, Mamin2014, Fang2013,Jamonneau2016}. This DQ Ramsey-based scheme can therefore disentangle magnetic and non-magnetic signals while also enabling improved, homogeneous per-pixel magnetic sensitivity across an image.

Previously, DQ Ramsey magnetic imaging has been hindered by the technical challenge of producing sufficiently uniform and strong MW fields to avoid spatially-varying errors in the optimal MW pulses and hence the NV$^\text{-}$ measurement protocol. Such pulse errors result in residual SQ coherence that remains sensitive to common-mode shifts of the $\ket{\pm1}$ sublevels, degrading the robustness of DQ magnetometry to stress-induced shifts and temperature drifts. 

The present work circumvents this challenge with a DQ 4-Ramsey protocol specifically designed to suppress the contribution of residual SQ coherence. By properly selecting the spin-1 rotations applied in four consecutive Ramsey measurements (4-Ramsey), the DQ signal from each Ramsey measurement is preserved while the residual SQ signals cancel. This scheme is broadly applicable to both NV$^\text{-}$ ensemble imaging and bulk sensing modalities, simultaneously mitigating the pernicious effects of stress-gradients and temperature-induced drifts. Since the 4-Ramsey protocol is a straightforward extension of established phase-alternation schemes, implementation in an existing system does not typically require additional MW components.

After describing the NV$^\text{-}$ center and experimental apparatus in Sec.~II, we outline and experimentally demonstrate the DQ 4-Ramsey protocol (Sec.~III). In Sec.~IV, we use SQ and DQ Ramsey fringe imaging to characterize, pixel by pixel, the reduced spatial variation in $T_2^*$ and NV$^\text{-}$ resonance frequency when using the DQ sensing basis. Using the same field of view as in Sec.~IV, we then measure a $1.5\times$ improved median per-pixel sensitivity and a $4.7\times$ narrower spatial distribution of per-pixel sensitivity using the DQ sensing basis compared to the SQ basis (Sec.~V). In Sec.~VI we highlight next steps to further improve DC magnetic sensitivity and temporal resolution, and we provide an outlook describing envisioned applications for high-sensitivity, broadband magnetic microscopy using the DQ 4-Ramsey protocol.

\section{Experimental Methods}

The NV center is a $\text{C}_{\text{3v}}$ symmetric color center in diamond formed by substitution of a nitrogen atom adjacent to a vacancy in the carbon lattice. We restrict attention to the negatively charged NV$^\text{-}$ center, which has an electronic spin-triplet ($S=1$) ground state with a zero-field-splitting at room temperature $D \approx 2.87$~GHz between the $|m_s=0\rangle$ and $|m_s=\pm1\rangle$ magnetic sublevels as shown in Fig.~\ref{fig:fig1}(a). Application of an external magnetic field splits the $|\pm1\rangle$ sublevels by the Zeeman effect. In the presence of a magnetic field $\vec{B}$ exceeding $\approx$\SI{1}{\milli\tesla} aligned with the NV$^\text{-}$ symmetry axis $z$, the NV$^\text{-}$ ground-state Hamiltonian can be approximated as~\cite{Glenn2017,StrainPaper,marcusStrainHam,galiSpinStrain,maletinskyStrainTerms}:
\begin{equation}
H/h \approx [D(T)+M_z]S^2_z + \frac{\gamma}{2\pi} B_z S_z,
\label{NVham}
\end{equation}
where $S_z$ is the dimensionless spin-1 operator, $M_z$ is the axial spin-stress coupling parameter, $D(T)$ is the temperature-dependent zero-field-splitting, $B_z$ is the projection of the external magnetic field $\vec{B}$ along the NV$^\text{-}$ symmetry axis, and $\gamma /(2\pi)=$\SI{28.03}{\giga\hertz \raiseto{-1} \tesla} is the NV$^\text{-}$ gyromagnetic ratio. Transverse magnetic, electric, and crystal stress terms are neglected as motivated in Refs.~\cite{StrainPaper, P1DQ, Dolde2011} (see the Supplemental Material~\cite{suppl} for further discussion of the crystal stress terms). Under these assumptions, the observed spatial variations in NV$^\text{-}$ resonance frequencies and linewidths are attributed to axial stress gradients arising from stress inhomogeneity in the host diamond crystal. Note that for DQ coherence magnetometry, the relative phase accumulated between the $\ket{\pm1}$ sublevels is not only immune to common-mode energy level shifts (proportional to $S_z^2$ in Eq.~\ref{NVham}) but also doubly sensitive to magnetic fields~\cite{Fang2013, Mamin2014, P1DQ}.

The present study employs a QDM to image spin-state-dependent fluorescence from a 1-$\upmu$-thick nitrogen-doped CVD diamond layer ([$\text{N}_\text{total}$]$\,\approx\!20\,$ppm,  $^{12}\text{C}\!=\!99.995$\%, natural abundance nitrogen) grown by Element Six Ltd. on a ($2\times2\times0.5$)$\,$mm$^3$ high purity diamond substrate. Post-growth treatment via electron irradiation and annealing increased the $\text{NV}^\text{-}$ concentration in the nitrogen-doped layer to $\approx\,$2$\,$ppm. The magnitude and distribution of stress inhomogeneity in the selected sample is representative of typical diamonds fabricated for NV-based magnetic imaging (see Refs.~\cite{StrainPaper, Friel2009} for additional examples).  

An approximately \SI{150}{\micro\meter}  by \SI{300}{\micro\meter} region of the NV layer is illuminated with \SI{1}{\watt} of \SI{532}{\nano\meter} laser light in a total internal reflection (TIR) geometry [see Fig.~\ref{fig:fig1}(b)]; and the associated NV$^\text{-}$ fluorescence is collected onto either a Heliotis heliCam C3 camera or a Hamamatsu C10508 avalanche photodiode. The heliCam operates by subtracting alternate exposures in analog and then digitizing the resultant background-subtracted signal. This procedure enables the detected magnetic-field-dependent NV$^\text{-}$ fluorescence to fill the 10-bit dynamic range of each pixel for modulated magnetometry sequences synchronized with the camera exposures. With an external frame-rate of up to \SI{3.8}{\kilo\hertz}, the heliCam provides submillisecond temporal resolution; while the internal exposure rate of up to \SI{1}{\mega\hertz} enables the accumulation of signal from multiple Ramsey measurements, each a few microseconds in duration, per external frame (Supplemental Material~\cite{suppl}). Two signal generators with phase control synthesize the dual-tone MW fields required for DQ coherence magnetometry in the presence of a bias magnetic field (Appendix A). Control over the relative phase between the two MW tones enables selective coupling to different DQ superposition states as described in the following section. These MW fields are applied to the NV$^\text{-}$ ensemble using a millimeter-scale shorted coaxial loop. Figure~\ref{fig:fig1}(c) depicts the typical spatial variation in Rabi frequency.

\begin{figure*}[ht]
  \includegraphics[width=17cm]{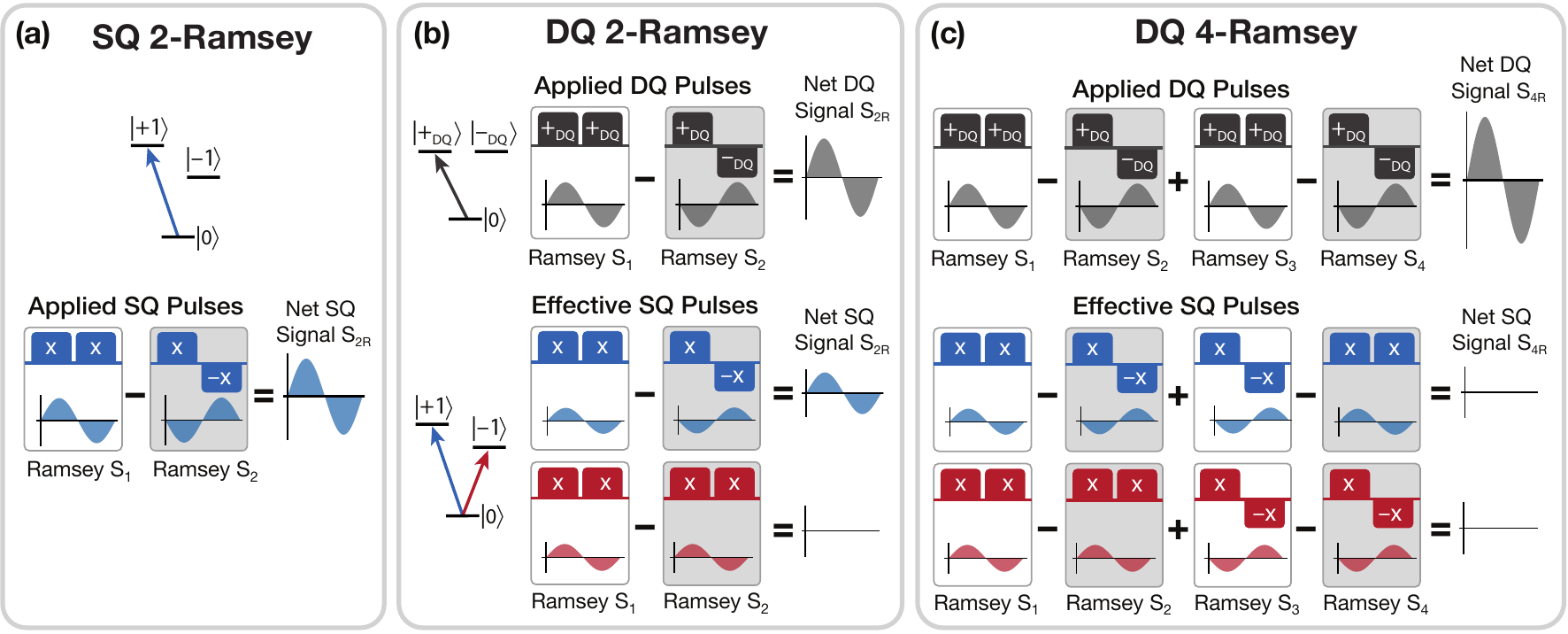} 
  \caption{\textbf{2-Ramsey and 4-Ramsey Measurement Protocols. (a)} 
    Representation of the single quantum (SQ) 2-Ramsey protocol. The choice of phases and resultant DC magnetometry signals are shown for each Ramsey sequence. The single-tone MW pulses address only the $\ket{0}\rightarrow\ket{+1}$ spin resonance.
    \textbf{(b)} Representation of the double quantum (DQ) 2-Ramsey protocol. In the top row (gray), the two-tone DQ pulses applied during each Ramsey sequence are depicted above the DC magnetometry curve associated with that choice of phases. The net DQ magnetometry signal S$_\text{2R}$ is shown on the right. In the middle and bottom rows, the applied MW pulses are decomposed into effective SQ rotations for each pseudo-two-level system. 
    Note that the SQ signals produced by the effective pulses addressing the $\ket{0}\rightarrow\ket{+1}$ transition (blue) do not cancel. As a result, SQ signals corrupt the resultant DQ 2-Ramsey signal when residual SQ coherence is present.
    \textbf{(c)} Representation of the DQ 4-Ramsey measurement protocol to cancel residual SQ signals resulting from MW pulse errors. The net DQ magnetometry signal S$_\text{4R}$ is shown on the right. In the presence of pulse errors, the resultant SQ DC magnetometry signals for each Ramsey sequence are depicted and shown to produce no net SQ signal when combined according to Eq.~\ref{eq:s4r}.
   }
\label{fig:fig2} 
\end{figure*}

\begin{figure}[ht]
  \includegraphics[width=8.5cm]{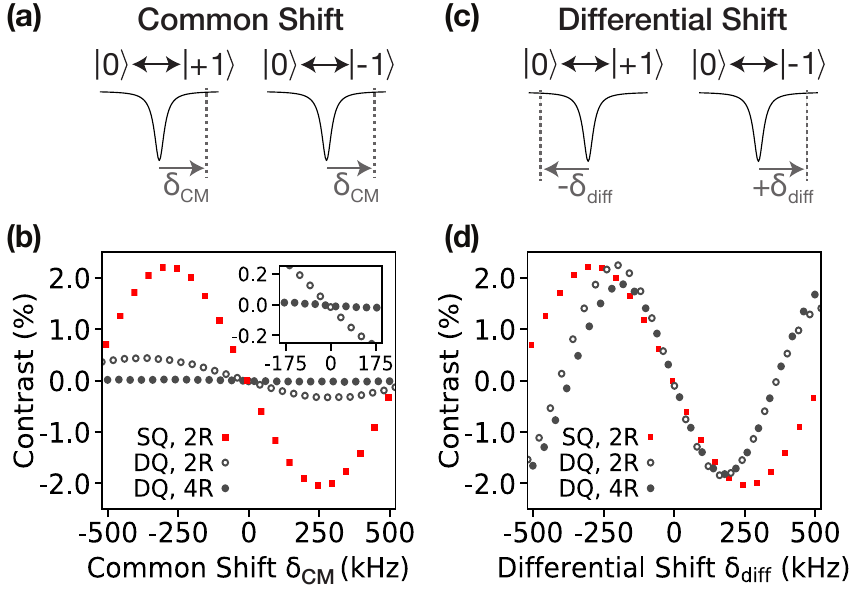} 
  \caption{\textbf{Measured Response to Common-Mode and Differential Detunings. (a)} Depiction of the applied MW field frequencies detuned from the NV$^\text{-}$ resonances in common mode by $\delta_\text{cm}$ to emulate stress- and temperature-induced shifts. 
  \textbf{(b)} Single-channel (photodiode) measurements of the NV$^\text{-}$ response to common-mode shifts of the $\ket{0}\rightarrow\ket{-1}$ and $\ket{0}\rightarrow\ket{+1}$ spin resonances. For each sensing protocol, $\delta_\text{cm}=0$ indicates the point of maximum slope after calibration (see Appendix~\ref{appendixMagOpt}). The SQ 2-Ramsey response, addressing only the $\ket{0}\rightarrow\ket{+1}$ spin resonance, is included for reference. The DQ 4-Ramsey response to common-mode shifts is suppressed by 96$\times$ compared to the SQ 2-Ramsey shift response.
  \textbf{(c)} Depiction of the applied MW field frequencies detuned from the NV$^\text{-}$ resonances differentially by $\pm\delta_\text{diff}$ to emulate axial-magnetic-field-induced shifts.
  \textbf{(d)} Single-channel measurements of the NV$^\text{-}$ response to differential shifts of the $\ket{0}\rightarrow\ket{-1}$ and $\ket{0}\rightarrow\ket{+1}$ spin resonances. For each measurement protocol, $\delta_\text{diff}=0$ indicates the point of maximum slope after calibration, which determines the optimal magnetometer sensitivity. The SQ 2-Ramsey response, addressing only the $\ket{0}\rightarrow\ket{+1}$ spin resonance, is included for reference.
  }
\label{fig:fig3} 
\end{figure}

\section{DQ 4-Ramsey Measurement Protocol}

We introduce a measurement protocol consisting of four consecutive Ramsey sequences that, when combined, isolate the desired DQ magnetometry signal from residual SQ signal by modulating the MW pulse phases. SQ protocols commonly employ sets of two Ramsey sequences (2-Ramsey), alternating the phase of the final $\pi$/2 pulse in successive sequences by $180 ^\circ$, to modulate the NV$^\text{-}$ fluorescence and cancel low-frequency noise, such as $1/f$ noise~\cite{Bar-Gill2013}. In such a SQ 2-Ramsey protocol, the magnetometry signal alternately maps to positive and negative changes in NV$^\text{-}$ fluorescence, such that subtracting every second detection from the previous yields a rectified magnetometry signal [see Fig.~\ref{fig:fig2}(a)].

Analogous DQ 2-Ramsey protocols exist: two-tone MW pulses couple the $\ket{0}$ state to equal-amplitude superpositions of the $\ket{\pm1}$ states, with a phase relationship $(\ket{+1} + e^{i \Delta \phi} \ket{-1})/2$ determined by the relative phase $\Delta \phi$ between the two MW tones~\cite{Mamin2014}. By modulating $\Delta \phi \!=\! \{\ang{0}, \ang{180}\}$ between the tones in the final $\pi/2$ pulse, the $\ket{0}$ state can be alternately coupled to the orthogonal superposition states $\ket{\pm\text{DQ}} = (\ket{+1} \pm \ket{-1})/\sqrt{2}$. Figure~\ref{fig:fig2}(b) depicts a representative DQ 2-Ramsey protocol.

Although DQ 2-Ramsey protocols effectively cancel noise at frequencies below the phase modulation frequency, this does not disentangle the desired DQ signal from unwanted SQ signal arising from MW pulse errors. In NV$^\text{-}$ ensemble measurements, MW pulse errors commonly arise from spatial gradients in the Rabi frequency across an interrogated ensemble or field of view, see Fig.~\ref{fig:fig1}(c) for an example of the typical Rabi gradient for a millimeter-scale shorted coaxial loop. Although the spatial properties of the MW control field depend upon setup-specific MW synthesis and delivery approaches, the 4-Ramsey protocol universally relaxes requirements on MW-field uniformity. The hyperfine splitting of the NV$^\text{-}$ resonances and stress-induced NV$^\text{-}$ resonance shifts can also introduce MW pulse errors via the detuning-dependent effective Rabi frequency. In this work, errors induced by the hyperfine splitting (\SI{2.2}{\mega\hertz} between each of the $m_I=\{-1,0,+1\}$ $^{14}$N nuclear spin states) are comparable to the Rabi gradient of $\pm\,$200$\,$kHz and uniform across the field of view. In addition, the spatially-correlated Rabi frequency variations on the 1$-$10$\,\upmu$m length scales in Fig.~\ref{fig:fig1}(c) are attributed to stress-induced shifts on the order of hundreds of kilohertz (see Sec. IV and Supplemental Material~\cite{suppl}).

We now describe the phase alternation pattern used in the DQ 4-Ramsey protocol to isolate DQ magnetic signals and present an experimental demonstration using photodiode-based measurements. 
Figure~\ref{fig:fig2}(c) depicts the resulting DQ rotations applied in the \{$\ket{0}$, $\ket{-\text{DQ}}$, $\ket{+\text{DQ}}$\} basis for a particular implementation of the DQ 4-Ramsey protocol, where the choice of relative phases is restricted to \ang{0} or \ang{180} (generalized phase requirements can be found in the Supplemental Material~\cite{suppl}). While the initial pulse in each Ramsey sequence prepares the $\ket{+\text{DQ}}$ state, the final pulse alternately couples to the $\ket{+\text{DQ}}$ and $\ket{-\text{DQ}}$ states, similar to the DQ 2-Ramsey protocol. If the signal from each of the four measurements $i = 1\!-\!4$ is denoted by $S_i$ then the rectified DQ signal $S_{4R}$ is given by
\begin{equation}\label{eq:s4r}
    S_{4R} = S_1 - S_2 + S_3 - S_4
\end{equation}
where, as shown in Fig.~\ref{fig:fig2}(c), $S_2$ and $S_4$ contain DQ signals with opposite sign compared to $S_1$ and $S_3$. 
When implementing these DQ rotations, we have flexibility in choosing the absolute phases of each tone. For example, \{\ang{0}, \ang{0}\} and \{\ang{180}, \ang{180}\} both couple to $\ket{+\text{DQ}}$ while \{\ang{0}, \ang{180}\} and \{\ang{180}, \ang{0}\} couple to $\ket{-\text{DQ}}$. We leverage this degree of freedom to ensure that residual SQ signals are canceled by Eq.~\ref{eq:s4r}. The effective SQ pulses applied to each two-level subsystem transition ($\ket{0} \rightarrow \ket{+1}$ and $\ket{0} \rightarrow \ket{-1}$) are illustrated in Fig.~\ref{fig:fig2}(c) as Bloch sphere rotations about the axes $x$ and $-x$. 

If pulse errors arise, leading to residual SQ coherence, then the resultant SQ signal contained in the summation $S_2$+$S_4$ is the same as $S_1$+$S_3$ (so long as the errors are constant over the $\sim 10\,\mu$s measurement duration). By subtracting these summations, $S_\text{4R}$ from Eq.~\ref{eq:s4r} eliminates this spurious SQ signal. When using the heliCam, Eq.~\ref{eq:s4r} is physically implemented by the on-chip circuitry, which subtracts alternating exposures in analog before digitization. For photodiode-based measurements, which provide access to $S_{1-4}$ directly, the right hand side of Eq.~\ref{eq:s4r} can be divided by the sum of $S_{1-4}$ to cancel the effects of multiplicative noise sources such as laser intensity fluctuations. 

Figure~\ref{fig:fig3} illustrates the benefit of the DQ 4-Ramsey protocol over SQ and DQ 2-Ramsey protocols. The measured changes in contrast in response to differential (magnetic-field-like) and common-mode (temperature, axial-stress-like) shifts are compared when operating with a free precession interval $\tau$ and detuning from the center hyperfine resonance, optimized for magnetic sensitivity (see Appendix~\ref{appendixMagOpt}). For the data presented in Figs.~\ref{fig:fig3}, NV$^\text{-}$ fluorescence from the same field of view as shown in Fig.~\ref{fig:fig1}(c) is collected onto a photodiode while sweeping the applied MW tone(s). By approximating the change in fluorescence about the optimal detuning ($\delta_\text{cm}$\,$=$\,$\delta_\text{diff}$\,$=$\,$0$) using a linear fit, we find that DQ Ramsey measurements using the conventional 2-Ramsey protocol (with residual SQ signal) suppress the response to common-mode shifts $\delta_\text{cm}$ compared to SQ 2-Ramsey measurements by a factor of 7. Although this suppression factor depends on both the particular setup and diamond, the factor of 7 reported in this work is similar to that in Ref.~\cite{Fang2013} for a single NV$^\text{-}$, which also attributes the residual observed response to MW pulse imperfections. Meanwhile, under the same experimental conditions, the DQ 4-Ramsey protocol suppresses the common shift response by about a factor of 100 compared to SQ Ramsey measurements. The residual DQ 4-Ramsey protocol response to common-mode shifts, visible in the inset of Fig.~\ref{fig:fig3}(b), is attributed to experimental imperfections when manipulating the phase of the MW control pulses. Alternative hardware implementations (e.g., using an arbitrary waveform generator) could likely yield further suppression of the DQ 4-Ramsey protocol response to common-mode shifts.

As depicted in Fig.~\ref{fig:fig3}(d), the DQ 4-Ramsey and DQ 2-Ramsey responses exhibit about a cumulative 25\% increase in slope (and an associated improvement in magnetometer sensitivity) compared to the SQ 2-Ramsey response, after accounting for the increased effective gyromagnetic ratio in the DQ basis and the loss of DQ contrast due to pulse errors. When each Ramsey signal $S_i$ is accessible, the bandwidth of the 4-Ramsey measurement is approximately half the bandwidth of the 2-Ramsey measurement. However, there is no corresponding decrease in sensitivity because the acquired DQ magnetic signals add constructively across the 4-Ramsey protocol (see Supplemental Material~\cite{suppl}).

\section{Ramsey Fringe Imaging}

We employ SQ 2-Ramsey and DQ 4-Ramsey measurements to image the NV$^\text{-}$ ensemble spin properties relevant for DC magnetic field sensitivity across a \SI{125}{\micro\meter} by \SI{125}{\micro\meter} field of view. The photon-shot-noise-limited sensitivity of a Ramsey-based measurement $\eta_\text{ramsey}$ depends upon the NV$^\text{-}$ ensemble dephasing time $T_2^*$, the contrast $C$, and the average number of photons collected per measurement $N$~\cite{barry_sensitivity_2019}:
\begin{equation}
    \eta_\text{ramsey} = \frac{1}{\gamma_{}} \frac{1}{\Delta m} \frac{1}{C e^{-(\tau / T_2^*)^p}\sqrt{N}} \frac{\sqrt{\tau+t_{r,i}}}{\tau}
\end{equation}
where $\Delta m$ accounts for the difference between the $m_s$ states used for the sensing basis ($\Delta m\!=\!1,2$ for the SQ, DQ bases), $\tau$ is the free precession interval per measurement, $p$ describes the decay shape, and $t_{r,i}$ indicates the duration of time dedicated to readout and initialization per measurement. The optimal free precession interval is determined by the NV$^\text{-}$ ensemble dephasing time $T_2^*$, which is proportional to the inverse of the inhomogeneous linewidth $\Gamma$ ($T_2^*\!=\,$1/$\pi\Gamma$ assuming a Lorentzian lineshape). Axial stress gradients within a pixel degrade $\eta_\text{ramsey}$ by decreasing $T_2^*$; stress-induced resonance shifts across an image both worsen $\eta_\text{ramsey}$ by ensuring that the chosen MW frequency is sub-optimal for all but a subset of pixels and introduce spatially-varying, non-magnetic offsets in the Ramsey signal that can complicate data analysis~\cite{StrainPaper}.

We image the NV$^\text{-}$ ensemble spin properties by sweeping the free precession time in the SQ and DQ Ramsey sequences and fitting the fringes to a sum of oscillations with a common decay envelope:
\begin{equation}
    S_\text{ramsey}(\tau) = e^{-\tau/T_2^*} \sum_{i = m_I} A_i \sin (2\pi f_i+\delta_i)
    \label{eqn_ramseysig}
\end{equation}
where each oscillatory term, indexed by $m_I = \{-1,0,1\}$ (for an $^{14}\text{N}$ ensemble), has an amplitude $A_i$, frequency $f_i$, phase shift $\delta_i$, and decay shape fixed to $p=1$~\cite{P1DQ}. A purposeful detuning of $3\,$MHz from the resonance corresponding to the $m_I=0$ hyperfine population was introduced in order to more easily extract all three frequencies and the decay envelope. Eq.~\ref{eqn_ramseysig} was rapidly fit to the data pixel-by-pixel using the open source, GPU-accelerated non-linear least-squares fitting software, GPUfit \cite{GPUfit}. The typical 95\% confidence intervals (C.I.) for the extracted dephasing times $T_2^*$ and amplitudes $A_i$ discussed below are less than 5\%, while the typical C.I. for $f_i$ are about 0.5\%.

\begin{figure}[t]
  \includegraphics[width=8.5cm]{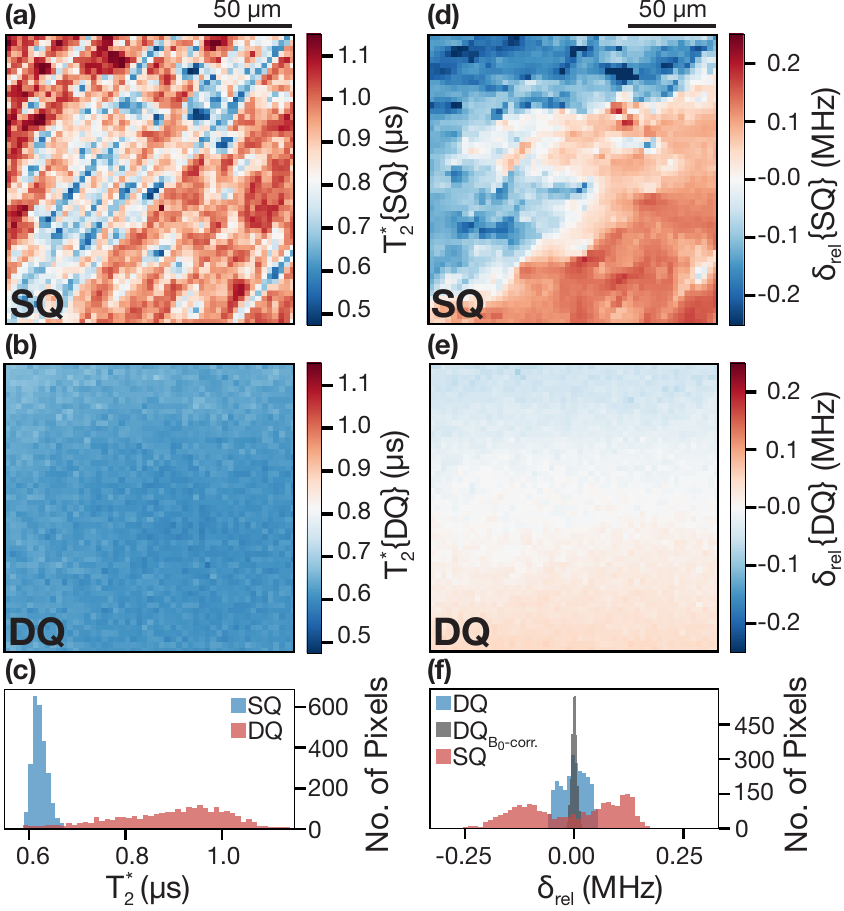} 
  \caption{\textbf{Imaging Ensemble Spin Properties. (a)} Image of the single quantum (SQ) $T_2^*$ extracted by fitting the SQ 2-Ramsey fringe decay to Eq.~\ref{eqn_ramseysig}. The field of view is \SI{125}{\micro\meter} by \SI{125}{\micro\meter}. Spatial variations in $T_2^*$\{\text{SQ}\} are due to stress-induced broadening of the NV$^\text{-}$ resonances within the three-dimensional volume imaged onto a pixel. 
  \textbf{(b)} Image of the double quantum $T_2^*$\{DQ\} measured using the DQ 4-Ramsey protocol across the same field of view as shown in (a). In pixels with minimal stress gradients, the $T_2^*$\{DQ\} is half the $T_2^*$\{SQ\}, as expected, due to the effectively doubled dipolar coupling to the surrounding paramagnetic spin bath, which dominates the NV$^\text{-}$ dephasing~\cite{P1DQ}.
   \textbf{(c)} Histogram of $T_2^*$\{SQ\} and $T_2^*$\{DQ\} values from the pixels in (a) and (b).
  \textbf{(d)} Image of the relative SQ resonance shifts $\delta_\text{rel}$\{SQ\} from the median SQ Ramsey fringe frequency. Variations in $\delta_\text{rel}$\{SQ\} are attributed predominantly to axial-stress-induced shifts of the NV$^\text{-}$ resonance frequencies between pixels.
  \textbf{(e)} Image of the relative DQ detuning $\delta_\text{rel}$\{DQ\} across the same field of view as shown in (a, b, d). The axial-stress-induced shifts apparent in (d) are mitigated. Inhomogeneity in the magnitude of the applied bias magnetic field $B_0$ results in a residual gradient of less than 1.4$\,\upmu$m.
  \textbf{(f)} Histogram of the extracted SQ and DQ $\delta_\text{rel}$ values from the pixels in (d) and (e). The distribution of DQ $\delta_\text{rel}$ values with the setup-specific $B_0$-gradient contribution corrected is shown in gray.
  }
\label{fig:fig4}
\end{figure}

\textit{Dephasing times} -- The extracted $T_2^*$ values for the SQ and DQ sensing bases are shown as images in Fig.~\ref{fig:fig4}(a) and \ref{fig:fig4}(b) and plotted as a histogram in Fig.~\ref{fig:fig4}(c). To quantify the non-normal spread in $T_2^*$ values, we report the median value and the relative inter-decile range (RIDR):
\begin{equation}\label{eqn:RIDR}
    \sigma_{\text{RIDR}} = \frac{D_{90}-D_{10}}{\text{(median)}}
\end{equation}
where 80\% of the measured values fall between the first decile $D_{10}$ and ninth decile $D_{90}$. In Fig.~\ref{fig:fig4}(a), the extracted $T_{2}^*\{\text{SQ}\}$ values have a median of 0.907$\,$(0.710, 1.03)$\,\upmu$s, where the values in parentheses correspond to the deciles (D$_{10}$,  D$_{90}$). As shown in Table~\RNC{1}, the calculated RIDR for the extracted $T_{2}^*\{\text{SQ}\}$ values is 35\% We attribute the spatially-correlated variations in $T_{2}^*\{\text{SQ}\}$ to axial stress gradients within pixels~\cite{StrainPaper,P1DQ}. The observed stress features are likely due to polishing-induced imperfections in the substrate surface upon which the NV$^\text{-}$ ensemble layer was grown~\cite{Friel2009}. 

Invulnerable to within-pixel stress gradients, the measured $T_{2}^*\{\text{DQ}\}$ values are 5.6$\times$ more uniform than the $T_{2}^*\{\text{SQ}\}$ values with a median of 0.621\,(0.605, 0.643)$\,\upmu$s and an RIDR of 6.0\%. Additionally, the median $T_{2}^*\{\text{DQ}\}$ is approximately one half the longest measured $T_{2}^*\{\text{SQ}\}$, \SI{1.15(3)}{\micro\second}, as expected when stress-induced dephasing is negligible and the dominant contribution to $T_2^*$ is dipolar coupling to an electronic spin bath (in this case of predominantly neutral substitutional nitrogen)~\cite{P1DQ}.

\begin{table*}[t]
\centering
\begin{tabular}{lcccc}
\hline \hline
 & \multicolumn{2}{c}{SQ}  & \multicolumn{2}{c}{DQ}        \\
\cline{2-3}
& $\tilde{x}$ (D$_\text{10}$, D$_\text{90}$) & RIDR & $\tilde{x}$  (D$_\text{10}$, D$_\text{90}$) & RIDR \\
\hline
Dephasing Time, $T_2^*\,$(\SI{}{\micro \second}) & 0.907 (0.710, 1.03) & 35\%  & 0.621 (0.605, 0.643) & 6.0\%  \\
Fringe Freq., $f_0$ $\,$(MHz)                    & 3.09 (2.94, 3.22)   & -- & 6.00 (5.96, 6.04)    & --  \\
Fringe Freq., $f_0$ ($B_0$-corr.)$\,$(MHz)       & 3.10 (2.95, 3.21)   & -- & 6.00 (5.99, 6.01)    & -- \\
Fringe Amplitude, $A_0$ $\,$(D.U.)               & 72.1 (61.8, 76.6)   & 21\%  & 73.5 (66.5, 77.0)  & 14\%  \\
\hline
\end{tabular}
\caption{Median extracted fit parameters ($\tilde{x}$) using Eq.~\ref{eqn_ramseysig} for SQ and DQ Ramsey fringe imaging. The lower and upper deciles, D$_\text{10}$ and D$_\text{10}$ are given in parentheses (80\% of the pixels exhibit values between D$_\text{10}$ and D$_\text{90}$). The SQ and DQ relative inter-decile ranges (RIDR) calculated using Eq.~\ref{eqn:RIDR} are included.}
\label{tab:table1}
\end{table*}


\textit{Fringe frequencies} -- Figures~\ref{fig:fig4}(d)- \ref{fig:fig4}(f) display the extracted SQ and DQ Ramsey fringe frequencies associated with the detuning of the applied MW pulses from the spin transition frequency for the $m_I\!=\!0$ hyperfine population. The relative detuning $\delta_\text{rel}$ from the median Ramsey fringe frequency $f_0$ is shown in Figs.~\ref{fig:fig4}(d)- \ref{fig:fig4}(f) to highlight the inhomogeneity across the field of view. The median SQ fringe frequency, $f_0\{\text{SQ}\}$, is 3.09\,(2.94, 3.22)\,MHz. The absolute spread in $f_0\{\text{SQ}\}$, $|D_{90}-D_{10}|\!=\,$280\,(14)\,kHz, is comparable to the median NV$^\text{-}$ resonance linewidth and attributed to stress gradients spanning multiple pixels~\cite{StrainPaper}.

\textit{Contrast} -- In the present work, inhomogeneity in the measurement contrast $C$ is largely independent of the choice of sensing basis (SQ or DQ) and is attributed to the Gaussian intensity profile of the excitation beam and fixed exposure duration. The extracted amplitudes $A_i$ for the measured Ramsey fringes, which are proportional to $C$, are reported in digital units (D.U.) of accumulated difference as measured by the heliCam C3. The median amplitudes $A_0\{\text{SQ}\}$ and $A_0\{\text{DQ}\}$ [$72.1\,(61.8,\,76.6)\,$D.U. and $73.5\,(66.5,\,77.0)\,$D.U.] as well as the RIDR ($21$\% and $14$\%) are comparable and included in Table~\ref{tab:table1}. 
Images of $A_0\{\text{SQ}\}$ and $A_0\{\text{DQ}\}$ are provided in the Supplemental Material~\cite{suppl} for reference.

\section{Magnetic Sensitivity Analysis}

We now compare the magnetic sensitivity of the SQ 2-Ramsey and DQ 4-Ramsey protocols across the same field of view described in Sec.~\RNC{4}. The narrower distribution of $T_2^*$\{DQ\} and resonance shifts $\delta_\text{rel}$\{DQ\} translate into improved, more homogeneous magnetic sensitivity. For both sensing bases, we select an optimal free precession interval $\tau$ and applied MW frequency (or frequencies) $f_\text{mw}$ to maximize the median NV$^\text{-}$ response to a change in magnetic field, $dS/dB$ (see Appendix~\ref{appendixMagOpt}). Under these conditions, a series of measurements is collected and used to determine the magnetic sensitivity pixel-by-pixel. 

The magnetic-field sensitivity is defined as $\delta B \sqrt{T_m}$, where $T_m$ is the measurement duration and $\delta B$ is the minimum detectable magnetic field, i.e., the field giving a signal-to-noise ratio (SNR) of 1~\cite{Taylor2008,LeSage2012,Bal2012,Schoenfeld2011}. A measurement with duration $T_m$ and sampling frequency $F_s = 1/T_m$ has a Nyquist-limited single-sided bandwidth of $\Delta f = F_s/2$. When the measurement bandwidth is sampling-rate limited, the minimum detectable magnetic field, $\delta B$ , is given by the standard deviation of a series of measurements, $\sigma_B$. The sensitivity to fields within that bandwidth $\Delta f$ can therefore be expressed as~\cite{SVMPaper}: 
\begin{equation}
    \eta = \sigma_B \sqrt{T_m} = \frac{\sigma_B}{\sqrt{2 \Delta f}}.
\end{equation}
In the present work, $F_s \approx 1.5\,$kHz, set by the camera's external frame rate. Each external frame contains the accumulated difference signal of multiple Ramsey sequences acquired at an internal exposure rate of approximately $140\,$kHz (see Supplemental Material ~\cite{suppl}). The standard deviation of each pixel was calculated from 1250 consecutive frames (\SI{1}{s} of acquired data) and converted to magnetic field units using the calibration $dS/dB$ measured for each pixel. Allan deviations of measurements using the SQ and DQ sensing bases are provided in the Supplemental Material~\cite{suppl}. Although the fixed time required to transfer data from the camera's 500-frame buffer ($\approx\,$5$\,$s, neglected in the above analysis) prevents continuous field monitoring at the calculated sensitivity for arbitrarily long times, the buffer still allows sets of high-bandwidth imaging data to be acquired over 0.4$\,$s.

\begin{figure}[t]
  \centering
  \includegraphics[width=8.5cm]{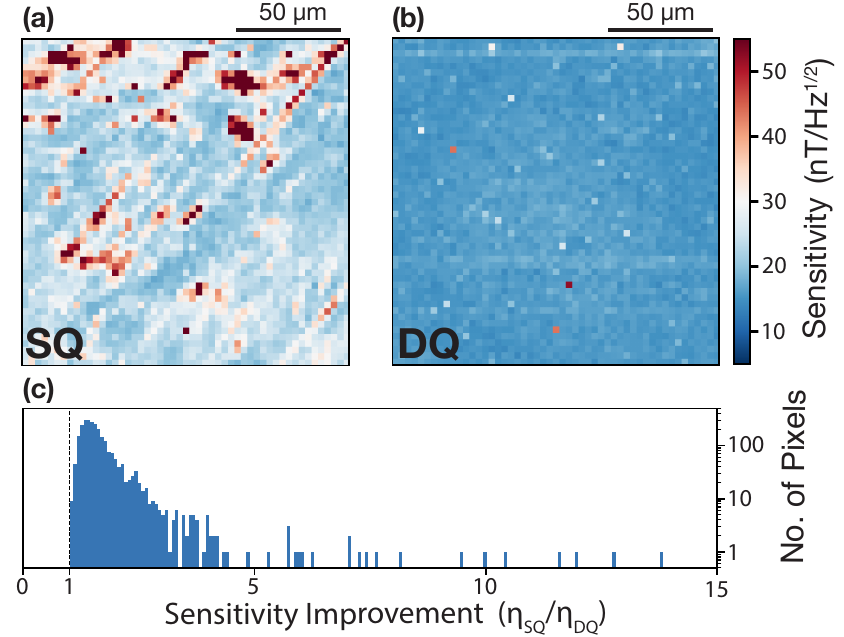} 
  \caption{\textbf{Imaging DC Magnetic Sensitivity. (a)} Data acquired with SQ 2-Ramsey protocol when operating at an applied MW field $f_\text{mw}$ and free precession interval $\tau$ that optimize the median per-pixel magnetic sensitivity. 
  \textbf{(b)} Data acquired with DQ 4-Ramsey protocol with optimal $f_{mw}'$ and $\tau'$, across the same field of view as (a). A few isolated, defective pixels with degraded sensitivity are visible.
  \textbf{(c)} Histogram of the improvement in relative sensitivity $\eta_{\text{SQ}}/\eta_{\text{DQ}}$ per pixel. 
  }
  \label{fig:fig5}
\end{figure}

The resulting sensitivities $\eta_\text{DQ}$ and $\eta_\text{SQ}$ are plotted in Figs.~\ref{fig:fig5}(a) and \ref{fig:fig5}(b). The median DQ 4-Ramsey per-pixel magnetic sensitivity $\eta_\text{DQ}\!=\,$15\,(14,\,16)\,nT\,Hz$^{-1/2}$ provides a factor of about 1.5$\times$ improvement compared to the SQ 2-Ramsey per-pixel magnetic sensitivity, $\eta_\text{SQ}\!=\,$22\,(19,\,34)\,nT\,Hz$^{-1/2}$ with voxel dimensions of ($2.5\times2.4\times1$)$\,$\SI{}{\raiseto{3}\micro\meter}. The upper and lower deciles, D$_{10}$ and D$_{90}$, are reported in parentheses. The typical uncertainty in the calculated per-pixel magnetic sensitivity, about 6\,\%, is dominated by the uncertainty in determining the parameters extracted from fitting the DC magnetometry curve in each pixel. 

The median ($D_{10}$, $D_{90}$) volume-normalized sensitivities are therefore $\eta^\text{V}_\text{DQ}\!=\,$34\,(32,\,37)\,nT\,Hz$^{-1/2}\,\upmu$m$^{3/2}$ and $\eta^\text{V}_\text{SQ}\!=\,$53\,(44,\,79)\,nT\,Hz$^{-1/2}\,\upmu$m$^{3/2}$. 
We observe about a 4.7$\times$ reduction in the RIDR for $\eta_\text{DQ}$ ($\approx14$\%) compared to the RIDR of $\eta_\text{SQ}$ ($\approx67$\%). The improved median sensitivity and reduced spread across the field of view are attributed to the elimination of axial-stress-induced dephasing and resonance shifts for the DQ 4-Ramsey protocol, such that it is possible to operate at the optimal $\tau$ and applied MW frequencies $f_\text{mw}$ for an increased fraction of pixels simultaneously. 

As illustrated in Fig.~\ref{fig:fig5}(c), all pixels exhibit improved magnetic sensitivity in the DQ sensing basis. Order-of-magnitude sensitivity improvements in the DQ basis are seen for the pixels corresponding to regions of diamond with higher stress gradients. In pixels with minimal stress-related effects, the improved magnetic sensitivity is attributed to (a) values of $f_\text{mw}$ and $\tau$ that are more optimal for an increased fraction of the pixels (see Appendix~\ref{appendixMagOpt}) and (b) the effectively doubled gyromagnetic ratio in the DQ sensing basis. The latter enables faster measurements (increased maximum $F_s$ because $T_2^*\{\text{DQ}\}$, and thus the optimal free precession interval $\tau$, is reduced compared to $T_2^*\{\text{SQ}\}$) for the same phase accumulation (see Supplemental Material~\cite{suppl}). The residual $14$\% spread in $\eta_\text{DQ}$ is a consequence of the Gaussian intensity profile of the excitation laser beam spot, which highlights the potential utility of optical beam-shaping techniques to enable further improvements. 

The median volume-normalized magnetic sensitivity $\eta^\text{V}_\text{DQ}\!=\,$34\,(32,\,37)\,nT\,Hz$^{-1/2}\,\upmu$m$^{3/2}$ demonstrated in this work coincidentally matches the value of \SI{34}{\nano\tesla \raiseto{3/2}\micro\meter \raiseto{-1/2}\hertz} estimated in Ref.~\cite{NeuronPaper}, which used photodiode-based CW-ODMR measurements to detect the single-neuron action potential from a living marine worm, \textit{M. infundibulum}. Critically, the present work achieves a similar sensitivity while operating in an imaging modality, with degraded optical collection efficiency, and using NV$^\text{-}$ centers along only a single crystal axis; whereas the non-imaging apparatus employed in Ref.~\cite{NeuronPaper} overlapped the resonances from two NV$^\text{-}$ axes and had an approximately $16\times$ higher optical collection efficiency.

\section{Outlook}

The demonstrated magnetic imaging method using the DQ 4-Ramsey protocol enables uniform magnetic sensitivity across a field of view independent of inhomogeneity in the host diamond material and applied microwave control fields. In particular, the MW phase alternation scheme of the 4-Ramsey protocol [Fig.~\ref{fig:fig2}(c)] isolates the double quantum magnetic signal from residual single quantum signal, decoupling the measurement from common-mode resonance frequency shifts induced by axial stress and temperature drift. The achieved 100$\times$ reduction in sensitivity to common-mode shifts is broadly advantageous, not only for magnetic imaging but also for single-channel applications such as magnetic navigation~\cite{canciani_absolute_2016}.

These methods provide a path toward imaging a range of dynamic magnetic phenomena, including nanotesla-scale fields from single mammalian neurons or cardiomyocytes, as well as fields from integrated circuits and condensed matter systems. Increased optical excitation intensity and further diamond material development could yield additional improvements in volume-normalized magnetic sensitivity. Although pulsed magnetometry protocols favor operating near the NV$^\text{-}$ center's saturation intensity (1-3\SI{}{\milli \watt \raiseto{-2} \micro \meter}~\cite{Wee2007}) to minimize the initialization and readout durations~\cite{barry_sensitivity_2019}, this work achieved optimal sensitivity when operating at an average intensity $\sim\!45\times$ below saturation. The lower intensity allowed the NV ensemble to maintain a favorable charge state fraction by reducing optical ionization of NV$^\text{-}$ to NV$^\text{0}$~\cite{alsid_photoluminescence_2019, aude_craik_microwave-assisted_2018}. For this reason, future material development improving and stabilizing the NV charge fraction, for example by reducing the density of other parasitic defects that can act as charge acceptors~\cite{purpleDiamond}, is critical.

The high-sensitivity, pulsed imaging method demonstrated here also enables applications beyond broadband magnetic microscopy such as parallelized, high-resolution NV$^\text{-}$ ensemble NMR using AC magnetic field detection protocols. 
Additionally, the MW phase control utilized for the DQ 4-Ramsey protocol is sufficient to implement magnetically-insensitive measurement protocols~\cite{Toyli2013, Hodges} as recently suggested by Ref.~\cite{Rajendran2017a} for imaging the lattice damage induced by colliding weakly-interacting massive particles (WIMPs). 

\section*{Acknowledgements}

We thank John Barry for early efforts evaluating the heliCam C3 camera and technical insights, Heliotis AG for experimental assistance implementing the heliCam C3 camera; Pauli Kehayias, David Phillips, Wilbur Lew, Arul Manickam, Jeff Cammerata, John Stetson, and Micheal DiMario for valuable discussions; and Kevin Olsson for feedback on the manuscript. This material is based upon work supported by, or in part by, the U.S. Army Research Laboratory and the U.S. Army Research Office under Grant No. W911NF-15-1-0548; the Army Research Laboratory MAQP program under Contract No. W911NF-19-2-0181; the DARPA DRINQS program (Grant No. D18AC00033); the National Science Foundation (NSF) Physics of Living Systems (PoLS) program under Grant No. PHY-1504610; the Air Force Office of Scientific Research Award No. FA9550-17-1-0371; the Department of Energy (DOE) Quantum Information Science Enabled Discovery (QuantISED) program under Award No. DE-SC0019396; Lockheed Martin under Contract No. A32198; and the University of Maryland Quantum Technology Center. J.M.S. was supported  by a Fannie and John Hertz Foundation Graduate Fellowship and a National Science Foundation Graduate Research Fellowship under Grant No. 1122374. 

\section*{Author contributions statement}

C.A.H., J.M.S., M.J.T., and E.B. conceived the experiments. C.A.H. and P.J.S. conducted the experiments. C.A.H. and J.M.S. analyzed the results. All authors contributed to and reviewed the manuscript. R.W. supervised the work.

\appendix

\section{Experimental Details}

An Agilent E9310A with built-in IQ modulation and a Windfreak SynthHD signal generator in combination with an external Marki-1545LMP IQ mixer provides the two-tone MW control fields and requisite phase control employed in this work. A Pulseblaster ESR-Pro with a 500$\,$MHz clock controls the synchronization of applied MW pulses, optical pulses, and camera exposures (or photodiode readouts when applicable). Samarium cobalt ring-shaped magnets (as described in~\cite{P1DQ}) apply a \SI{5}{\milli\tesla} bias magnetic field used to split the $\ket{0}$ to $\ket{\pm1}$ transitions.

\section{NV$^\text{-}$ Ensemble Magnetometer Calibration}
\label{appendixMagOpt}

For the measurements in this work, the optimal free precession interval $\tau$ and applied MW frequency $f_\text{mw}$ are chosen to maximize the NV$^\text{-}$ response $S$ to changes in magnetic-field $\frac{dS}{dB}$ (i.e., minimize the sensitivity $\eta$). Although the optimal $\tau$ is approximately equal to $T_2^*$~\cite{barry_sensitivity_2019}, the Ramsey fringe beating introduced by the hyperfine splitting of the NV$^\text{-}$ restricts the possible choices of $\tau$ to discrete values. As a consequence, we select the nearest available $\tau$ to $T_2^*$ for each sensing basis. With the free precession interval $\tau$ fixed, the optimal $f_\text{mw}$ is determined by sweeping the applied MW frequency to emulate a change in magnetic field, producing a DC magnetometry curve from which $f_\text{mw}$ is chosen to maximize the slope $\frac{dS}{dB}$. To determine the optimal MW frequencies for DQ Ramsey measurements, $f_\text{mw}'$, the two applied MW tones are swept differentially (one tone with positive detuning $+\delta$ and the second tone with an equal but opposite detuning $-\delta$). As with the SQ calibration, the values of $f_\text{mw}'$ are chosen to maximize the NV$^\text{-}$ response $\frac{dS}{dB}$. For all measurements using the heliCam, the free precession interval and MW frequency (or frequencies) are chosen to minimize the median per-pixel sensitivity across the field of view.

\bibliography{references}

\begin{thebibliography}{41}%
\makeatletter
\providecommand \@ifxundefined [1]{%
 \@ifx{#1\undefined}
}%
\providecommand \@ifnum [1]{%
 \ifnum #1\expandafter \@firstoftwo
 \else \expandafter \@secondoftwo
 \fi
}%
\providecommand \@ifx [1]{%
 \ifx #1\expandafter \@firstoftwo
 \else \expandafter \@secondoftwo
 \fi
}%
\providecommand \natexlab [1]{#1}%
\providecommand \enquote  [1]{``#1''}%
\providecommand \bibnamefont  [1]{#1}%
\providecommand \bibfnamefont [1]{#1}%
\providecommand \citenamefont [1]{#1}%
\providecommand \href@noop [0]{\@secondoftwo}%
\providecommand \href [0]{\begingroup \@sanitize@url \@href}%
\providecommand \@href[1]{\@@startlink{#1}\@@href}%
\providecommand \@@href[1]{\endgroup#1\@@endlink}%
\providecommand \@sanitize@url [0]{\catcode `\\12\catcode `\$12\catcode
  `\&12\catcode `\#12\catcode `\^12\catcode `\_12\catcode `\%12\relax}%
\providecommand \@@startlink[1]{}%
\providecommand \@@endlink[0]{}%
\providecommand \url  [0]{\begingroup\@sanitize@url \@url }%
\providecommand \@url [1]{\endgroup\@href {#1}{\urlprefix }}%
\providecommand \urlprefix  [0]{URL }%
\providecommand \Eprint [0]{\href }%
\providecommand \doibase [0]{https://doi.org/}%
\providecommand \selectlanguage [0]{\@gobble}%
\providecommand \bibinfo  [0]{\@secondoftwo}%
\providecommand \bibfield  [0]{\@secondoftwo}%
\providecommand \translation [1]{[#1]}%
\providecommand \BibitemOpen [0]{}%
\providecommand \bibitemStop [0]{}%
\providecommand \bibitemNoStop [0]{.\EOS\space}%
\providecommand \EOS [0]{\spacefactor3000\relax}%
\providecommand \BibitemShut  [1]{\csname bibitem#1\endcsname}%
\let\auto@bib@innerbib\@empty
\bibitem [{\citenamefont {Rondin}\ \emph {et~al.}(2014)\citenamefont {Rondin},
  \citenamefont {Tetienne}, \citenamefont {Hingant}, \citenamefont {Roch},
  \citenamefont {Maletinsky},\ and\ \citenamefont {Jacques}}]{Rondin2014}%
  \BibitemOpen
  \bibfield  {author} {\bibinfo {author} {\bibfnamefont {L.}~\bibnamefont
  {Rondin}}, \bibinfo {author} {\bibfnamefont {J.~P.}\ \bibnamefont
  {Tetienne}}, \bibinfo {author} {\bibfnamefont {T.}~\bibnamefont {Hingant}},
  \bibinfo {author} {\bibfnamefont {J.~F.}\ \bibnamefont {Roch}}, \bibinfo
  {author} {\bibfnamefont {P.}~\bibnamefont {Maletinsky}},\ and\ \bibinfo
  {author} {\bibfnamefont {V.}~\bibnamefont {Jacques}},\ }\bibfield  {title}
  {\bibinfo {title} {{Magnetometry with nitrogen-vacancy defects in diamond}},\
  }\href {https://doi.org/10.1088/0034-4885/77/5/056503} {\bibfield  {journal}
  {\bibinfo  {journal} {Reports on Progress in Physics}\ }\textbf {\bibinfo
  {volume} {77}},\ \bibinfo {pages} {056503} (\bibinfo {year}
  {2014})}\BibitemShut {NoStop}%
\bibitem [{\citenamefont {Doherty}\ \emph {et~al.}(2013)\citenamefont
  {Doherty}, \citenamefont {Manson}, \citenamefont {Delaney}, \citenamefont
  {Jelezko}, \citenamefont {Wrachtrup},\ and\ \citenamefont
  {Hollenberg}}]{Doherty2013}%
  \BibitemOpen
  \bibfield  {author} {\bibinfo {author} {\bibfnamefont {M.~W.}\ \bibnamefont
  {Doherty}}, \bibinfo {author} {\bibfnamefont {N.~B.}\ \bibnamefont {Manson}},
  \bibinfo {author} {\bibfnamefont {P.}~\bibnamefont {Delaney}}, \bibinfo
  {author} {\bibfnamefont {F.}~\bibnamefont {Jelezko}}, \bibinfo {author}
  {\bibfnamefont {J.}~\bibnamefont {Wrachtrup}},\ and\ \bibinfo {author}
  {\bibfnamefont {L.~C.~L.}\ \bibnamefont {Hollenberg}},\ }\bibfield  {title}
  {\bibinfo {title} {{The nitrogen-vacancy colour centre in diamond}},\ }\href
  {https://doi.org/10.1016/j.physrep.2013.02.001} {\bibfield  {journal}
  {\bibinfo  {journal} {Physics Reports}\ }\textbf {\bibinfo {volume} {528}},\
  \bibinfo {pages} {1} (\bibinfo {year} {2013})}\BibitemShut {NoStop}%
\bibitem [{\citenamefont {Barry}\ \emph {et~al.}(2020)\citenamefont {Barry},
  \citenamefont {Schloss}, \citenamefont {Bauch}, \citenamefont {Turner},
  \citenamefont {Hart}, \citenamefont {Pham},\ and\ \citenamefont
  {Walsworth}}]{barry_sensitivity_2019}%
  \BibitemOpen
  \bibfield  {author} {\bibinfo {author} {\bibfnamefont {J.~F.}\ \bibnamefont
  {Barry}}, \bibinfo {author} {\bibfnamefont {J.~M.}\ \bibnamefont {Schloss}},
  \bibinfo {author} {\bibfnamefont {E.}~\bibnamefont {Bauch}}, \bibinfo
  {author} {\bibfnamefont {M.~J.}\ \bibnamefont {Turner}}, \bibinfo {author}
  {\bibfnamefont {C.~A.}\ \bibnamefont {Hart}}, \bibinfo {author}
  {\bibfnamefont {L.~M.}\ \bibnamefont {Pham}},\ and\ \bibinfo {author}
  {\bibfnamefont {R.~L.}\ \bibnamefont {Walsworth}},\ }\bibfield  {title}
  {\bibinfo {title} {Sensitivity optimization for nv-diamond magnetometry},\
  }\href {https://doi.org/10.1103/RevModPhys.92.015004} {\bibfield  {journal}
  {\bibinfo  {journal} {Rev. Mod. Phys.}\ }\textbf {\bibinfo {volume} {92}},\
  \bibinfo {pages} {015004} (\bibinfo {year} {2020})}\BibitemShut {NoStop}%
\bibitem [{\citenamefont {Levine}\ \emph {et~al.}(2019)\citenamefont {Levine},
  \citenamefont {Turner}, \citenamefont {Kehayias}, \citenamefont {Hart},
  \citenamefont {Langellier}, \citenamefont {Trubko}, \citenamefont {Glenn},
  \citenamefont {Fu},\ and\ \citenamefont {Walsworth}}]{Levine2019}%
  \BibitemOpen
  \bibfield  {author} {\bibinfo {author} {\bibfnamefont {E.~V.}\ \bibnamefont
  {Levine}}, \bibinfo {author} {\bibfnamefont {M.~J.}\ \bibnamefont {Turner}},
  \bibinfo {author} {\bibfnamefont {P.}~\bibnamefont {Kehayias}}, \bibinfo
  {author} {\bibfnamefont {C.~A.}\ \bibnamefont {Hart}}, \bibinfo {author}
  {\bibfnamefont {N.}~\bibnamefont {Langellier}}, \bibinfo {author}
  {\bibfnamefont {R.}~\bibnamefont {Trubko}}, \bibinfo {author} {\bibfnamefont
  {D.~R.}\ \bibnamefont {Glenn}}, \bibinfo {author} {\bibfnamefont {R.~R.}\
  \bibnamefont {Fu}},\ and\ \bibinfo {author} {\bibfnamefont {R.~L.}\
  \bibnamefont {Walsworth}},\ }\bibfield  {title} {\bibinfo {title} {Principles
  and techniques of the quantum diamond microscope},\ }\href
  {https://doi.org/https://doi.org/10.1515/nanoph-2019-0209} {\bibfield
  {journal} {\bibinfo  {journal} {Nanophotonics}\ }\textbf {\bibinfo {volume}
  {8}},\ \bibinfo {pages} {1945 } (\bibinfo {year} {2019})}\BibitemShut
  {NoStop}%
\bibitem [{\citenamefont {Glenn}\ \emph {et~al.}(2017)\citenamefont {Glenn},
  \citenamefont {Fu}, \citenamefont {Kehayias}, \citenamefont {{Le Sage}},
  \citenamefont {Lima}, \citenamefont {Weiss},\ and\ \citenamefont
  {Walsworth}}]{Glenn2017}%
  \BibitemOpen
  \bibfield  {author} {\bibinfo {author} {\bibfnamefont {D.~R.}\ \bibnamefont
  {Glenn}}, \bibinfo {author} {\bibfnamefont {R.~R.}\ \bibnamefont {Fu}},
  \bibinfo {author} {\bibfnamefont {P.}~\bibnamefont {Kehayias}}, \bibinfo
  {author} {\bibfnamefont {D.}~\bibnamefont {{Le Sage}}}, \bibinfo {author}
  {\bibfnamefont {E.~A.}\ \bibnamefont {Lima}}, \bibinfo {author}
  {\bibfnamefont {B.~P.}\ \bibnamefont {Weiss}},\ and\ \bibinfo {author}
  {\bibfnamefont {R.~L.}\ \bibnamefont {Walsworth}},\ }\bibfield  {title}
  {\bibinfo {title} {{Micrometer‐scale magnetic imaging of geological samples
  using a quantum diamond microscope}},\ }\href
  {https://doi.org/10.1002/2017GC006946} {\bibfield  {journal} {\bibinfo
  {journal} {Geochemistry, Geophysics, Geosystems}\ }\textbf {\bibinfo {volume}
  {18}},\ \bibinfo {pages} {3254} (\bibinfo {year} {2017})}\BibitemShut
  {NoStop}%
\bibitem [{\citenamefont {Simpson}\ \emph {et~al.}(2016)\citenamefont
  {Simpson}, \citenamefont {Tetienne}, \citenamefont {McCoey}, \citenamefont
  {Ganesan}, \citenamefont {Hall}, \citenamefont {Petrou}, \citenamefont
  {Scholten},\ and\ \citenamefont {Hollenberg}}]{Simpson2016}%
  \BibitemOpen
  \bibfield  {author} {\bibinfo {author} {\bibfnamefont {D.~A.}\ \bibnamefont
  {Simpson}}, \bibinfo {author} {\bibfnamefont {J.-P.}\ \bibnamefont
  {Tetienne}}, \bibinfo {author} {\bibfnamefont {J.~M.}\ \bibnamefont
  {McCoey}}, \bibinfo {author} {\bibfnamefont {K.}~\bibnamefont {Ganesan}},
  \bibinfo {author} {\bibfnamefont {L.~T.}\ \bibnamefont {Hall}}, \bibinfo
  {author} {\bibfnamefont {S.}~\bibnamefont {Petrou}}, \bibinfo {author}
  {\bibfnamefont {R.~E.}\ \bibnamefont {Scholten}},\ and\ \bibinfo {author}
  {\bibfnamefont {L.~C.~L.}\ \bibnamefont {Hollenberg}},\ }\bibfield  {title}
  {\bibinfo {title} {{Magneto-optical imaging of thin magnetic films using
  spins in diamond}},\ }\href {https://doi.org/10.1038/srep22797} {\bibfield
  {journal} {\bibinfo  {journal} {Scientific Reports}\ }\textbf {\bibinfo
  {volume} {6}},\ \bibinfo {pages} {22797} (\bibinfo {year}
  {2016})}\BibitemShut {NoStop}%
\bibitem [{\citenamefont {McCoey}\ \emph {et~al.}(2020)\citenamefont {McCoey},
  \citenamefont {Matsuoka}, \citenamefont {de~Gille}, \citenamefont {Hall},
  \citenamefont {Shaw}, \citenamefont {Tetienne}, \citenamefont {Kisailus},
  \citenamefont {Hollenberg},\ and\ \citenamefont {Simpson}}]{McCoey2020}%
  \BibitemOpen
  \bibfield  {author} {\bibinfo {author} {\bibfnamefont {J.~M.}\ \bibnamefont
  {McCoey}}, \bibinfo {author} {\bibfnamefont {M.}~\bibnamefont {Matsuoka}},
  \bibinfo {author} {\bibfnamefont {R.~W.}\ \bibnamefont {de~Gille}}, \bibinfo
  {author} {\bibfnamefont {L.~T.}\ \bibnamefont {Hall}}, \bibinfo {author}
  {\bibfnamefont {J.~A.}\ \bibnamefont {Shaw}}, \bibinfo {author}
  {\bibfnamefont {J.}~\bibnamefont {Tetienne}}, \bibinfo {author}
  {\bibfnamefont {D.}~\bibnamefont {Kisailus}}, \bibinfo {author}
  {\bibfnamefont {L.~C.~L.}\ \bibnamefont {Hollenberg}},\ and\ \bibinfo
  {author} {\bibfnamefont {D.~A.}\ \bibnamefont {Simpson}},\ }\bibfield
  {title} {\bibinfo {title} {{Quantum Magnetic Imaging of Iron
  Biomineralization in Teeth of the Chiton
  {\textless}i{\textgreater}Acanthopleura
  hirtosa{\textless}/i{\textgreater}}},\ }\href
  {https://doi.org/10.1002/smtd.201900754} {\bibfield  {journal} {\bibinfo
  {journal} {Small Methods}\ }\textbf {\bibinfo {volume} {4}},\ \bibinfo
  {pages} {1900754} (\bibinfo {year} {2020})}\BibitemShut {NoStop}%
\bibitem [{\citenamefont {Tetienne}\ \emph {et~al.}(2017)\citenamefont
  {Tetienne}, \citenamefont {Dontschuk}, \citenamefont {Broadway},
  \citenamefont {Stacey}, \citenamefont {Simpson},\ and\ \citenamefont
  {Hollenberg}}]{Tetienne2016}%
  \BibitemOpen
  \bibfield  {author} {\bibinfo {author} {\bibfnamefont {J.-P.}\ \bibnamefont
  {Tetienne}}, \bibinfo {author} {\bibfnamefont {N.}~\bibnamefont {Dontschuk}},
  \bibinfo {author} {\bibfnamefont {D.~A.}\ \bibnamefont {Broadway}}, \bibinfo
  {author} {\bibfnamefont {A.}~\bibnamefont {Stacey}}, \bibinfo {author}
  {\bibfnamefont {D.~A.}\ \bibnamefont {Simpson}},\ and\ \bibinfo {author}
  {\bibfnamefont {L.~C.~L.}\ \bibnamefont {Hollenberg}},\ }\bibfield  {title}
  {\bibinfo {title} {{Quantum imaging of current flow in graphene}},\ }\href
  {https://doi.org/10.1126/sciadv.1602429} {\bibfield  {journal} {\bibinfo
  {journal} {Science Advances}\ }\textbf {\bibinfo {volume} {3}},\ \bibinfo
  {pages} {e1602429} (\bibinfo {year} {2017})}\BibitemShut {NoStop}%
\bibitem [{\citenamefont {Ku}\ \emph {et~al.}(2020)\citenamefont {Ku},
  \citenamefont {Zhou}, \citenamefont {Li}, \citenamefont {Shin}, \citenamefont
  {Shi}, \citenamefont {Burch}, \citenamefont {Anderson}, \citenamefont
  {Pierce}, \citenamefont {Xie}, \citenamefont {Hamo}, \citenamefont {Vool},
  \citenamefont {Zhang}, \citenamefont {Casola}, \citenamefont {Taniguchi},
  \citenamefont {Watanabe}, \citenamefont {Fogler}, \citenamefont {Kim},
  \citenamefont {Yacoby},\ and\ \citenamefont {Walsworth}}]{Ku2020}%
  \BibitemOpen
  \bibfield  {author} {\bibinfo {author} {\bibfnamefont {M.~J.~H.}\
  \bibnamefont {Ku}}, \bibinfo {author} {\bibfnamefont {T.~X.}\ \bibnamefont
  {Zhou}}, \bibinfo {author} {\bibfnamefont {Q.}~\bibnamefont {Li}}, \bibinfo
  {author} {\bibfnamefont {Y.~J.}\ \bibnamefont {Shin}}, \bibinfo {author}
  {\bibfnamefont {J.~K.}\ \bibnamefont {Shi}}, \bibinfo {author} {\bibfnamefont
  {C.}~\bibnamefont {Burch}}, \bibinfo {author} {\bibfnamefont {L.~E.}\
  \bibnamefont {Anderson}}, \bibinfo {author} {\bibfnamefont {A.~T.}\
  \bibnamefont {Pierce}}, \bibinfo {author} {\bibfnamefont {Y.}~\bibnamefont
  {Xie}}, \bibinfo {author} {\bibfnamefont {A.}~\bibnamefont {Hamo}}, \bibinfo
  {author} {\bibfnamefont {U.}~\bibnamefont {Vool}}, \bibinfo {author}
  {\bibfnamefont {H.}~\bibnamefont {Zhang}}, \bibinfo {author} {\bibfnamefont
  {F.}~\bibnamefont {Casola}}, \bibinfo {author} {\bibfnamefont
  {T.}~\bibnamefont {Taniguchi}}, \bibinfo {author} {\bibfnamefont
  {K.}~\bibnamefont {Watanabe}}, \bibinfo {author} {\bibfnamefont {M.~M.}\
  \bibnamefont {Fogler}}, \bibinfo {author} {\bibfnamefont {P.}~\bibnamefont
  {Kim}}, \bibinfo {author} {\bibfnamefont {A.}~\bibnamefont {Yacoby}},\ and\
  \bibinfo {author} {\bibfnamefont {R.~L.}\ \bibnamefont {Walsworth}},\
  }\bibfield  {title} {\bibinfo {title} {{Imaging viscous flow of the Dirac
  fluid in graphene}},\ }\href {https://doi.org/10.1038/s41586-020-2507-2}
  {\bibfield  {journal} {\bibinfo  {journal} {Nature}\ }\textbf {\bibinfo
  {volume} {583}},\ \bibinfo {pages} {537} (\bibinfo {year}
  {2020})}\BibitemShut {NoStop}%
\bibitem [{\citenamefont {Turner}\ \emph {et~al.}(2020)\citenamefont {Turner},
  \citenamefont {Langellier}, \citenamefont {Bainbridge}, \citenamefont
  {Walters}, \citenamefont {Meesala}, \citenamefont {Babinec}, \citenamefont
  {Kehayias}, \citenamefont {Yacoby}, \citenamefont {Hu}, \citenamefont
  {Lon{\v{c}}ar}, \citenamefont {Walsworth},\ and\ \citenamefont
  {Levine}}]{ICpaper}%
  \BibitemOpen
  \bibfield  {author} {\bibinfo {author} {\bibfnamefont {M.~J.}\ \bibnamefont
  {Turner}}, \bibinfo {author} {\bibfnamefont {N.}~\bibnamefont {Langellier}},
  \bibinfo {author} {\bibfnamefont {R.}~\bibnamefont {Bainbridge}}, \bibinfo
  {author} {\bibfnamefont {D.}~\bibnamefont {Walters}}, \bibinfo {author}
  {\bibfnamefont {S.}~\bibnamefont {Meesala}}, \bibinfo {author} {\bibfnamefont
  {T.~M.}\ \bibnamefont {Babinec}}, \bibinfo {author} {\bibfnamefont
  {P.}~\bibnamefont {Kehayias}}, \bibinfo {author} {\bibfnamefont
  {A.}~\bibnamefont {Yacoby}}, \bibinfo {author} {\bibfnamefont
  {E.}~\bibnamefont {Hu}}, \bibinfo {author} {\bibfnamefont {M.}~\bibnamefont
  {Lon{\v{c}}ar}}, \bibinfo {author} {\bibfnamefont {R.~L.}\ \bibnamefont
  {Walsworth}},\ and\ \bibinfo {author} {\bibfnamefont {E.~V.}\ \bibnamefont
  {Levine}},\ }\bibfield  {title} {\bibinfo {title} {{Magnetic Field
  Fingerprinting of Integrated-Circuit Activity with a Quantum Diamond
  Microscope}},\ }\href {https://doi.org/10.1103/PhysRevApplied.14.014097}
  {\bibfield  {journal} {\bibinfo  {journal} {Physical Review Applied}\
  }\textbf {\bibinfo {volume} {14}},\ \bibinfo {pages} {014097} (\bibinfo
  {year} {2020})}\BibitemShut {NoStop}%
\bibitem [{\citenamefont {{Le Sage}}\ \emph {et~al.}(2013)\citenamefont {{Le
  Sage}}, \citenamefont {Arai}, \citenamefont {Glenn}, \citenamefont
  {DeVience}, \citenamefont {Pham}, \citenamefont {Rahn-Lee}, \citenamefont
  {Lukin}, \citenamefont {Yacoby}, \citenamefont {Komeili},\ and\ \citenamefont
  {Walsworth}}]{LeSage2013}%
  \BibitemOpen
  \bibfield  {author} {\bibinfo {author} {\bibfnamefont {D.}~\bibnamefont {{Le
  Sage}}}, \bibinfo {author} {\bibfnamefont {K.}~\bibnamefont {Arai}}, \bibinfo
  {author} {\bibfnamefont {D.~R.}\ \bibnamefont {Glenn}}, \bibinfo {author}
  {\bibfnamefont {S.~J.}\ \bibnamefont {DeVience}}, \bibinfo {author}
  {\bibfnamefont {L.~M.}\ \bibnamefont {Pham}}, \bibinfo {author}
  {\bibfnamefont {L.}~\bibnamefont {Rahn-Lee}}, \bibinfo {author}
  {\bibfnamefont {M.~D.}\ \bibnamefont {Lukin}}, \bibinfo {author}
  {\bibfnamefont {A.}~\bibnamefont {Yacoby}}, \bibinfo {author} {\bibfnamefont
  {A.}~\bibnamefont {Komeili}},\ and\ \bibinfo {author} {\bibfnamefont
  {R.}~\bibnamefont {Walsworth}},\ }\bibfield  {title} {\bibinfo {title}
  {{Optical magnetic imaging of living cells.}},\ }\href
  {https://doi.org/10.1038/nature12072} {\bibfield  {journal} {\bibinfo
  {journal} {Nature}\ }\textbf {\bibinfo {volume} {496}},\ \bibinfo {pages}
  {486} (\bibinfo {year} {2013})}\BibitemShut {NoStop}%
\bibitem [{\citenamefont {Glenn}\ \emph {et~al.}(2015)\citenamefont {Glenn},
  \citenamefont {Lee}, \citenamefont {Park}, \citenamefont {Weissleder},
  \citenamefont {Yacoby}, \citenamefont {Lukin}, \citenamefont {Lee},
  \citenamefont {Walsworth},\ and\ \citenamefont {Connolly}}]{Glenn2015}%
  \BibitemOpen
  \bibfield  {author} {\bibinfo {author} {\bibfnamefont {D.~R.}\ \bibnamefont
  {Glenn}}, \bibinfo {author} {\bibfnamefont {K.}~\bibnamefont {Lee}}, \bibinfo
  {author} {\bibfnamefont {H.}~\bibnamefont {Park}}, \bibinfo {author}
  {\bibfnamefont {R.}~\bibnamefont {Weissleder}}, \bibinfo {author}
  {\bibfnamefont {A.}~\bibnamefont {Yacoby}}, \bibinfo {author} {\bibfnamefont
  {M.~D.}\ \bibnamefont {Lukin}}, \bibinfo {author} {\bibfnamefont
  {H.}~\bibnamefont {Lee}}, \bibinfo {author} {\bibfnamefont {R.~L.}\
  \bibnamefont {Walsworth}},\ and\ \bibinfo {author} {\bibfnamefont {C.~B.}\
  \bibnamefont {Connolly}},\ }\bibfield  {title} {\bibinfo {title}
  {{Single-cell magnetic imaging using a quantum diamond microscope}},\ }\href
  {https://doi.org/10.1038/nmeth.3449} {\bibfield  {journal} {\bibinfo
  {journal} {Nature Methods}\ }\textbf {\bibinfo {volume} {12}},\ \bibinfo
  {pages} {736} (\bibinfo {year} {2015})}\BibitemShut {NoStop}%
\bibitem [{\citenamefont {Kehayias}\ \emph {et~al.}(2019)\citenamefont
  {Kehayias}, \citenamefont {Turner}, \citenamefont {Trubko}, \citenamefont
  {Schloss}, \citenamefont {Hart}, \citenamefont {Wesson}, \citenamefont
  {Glenn},\ and\ \citenamefont {Walsworth}}]{StrainPaper}%
  \BibitemOpen
  \bibfield  {author} {\bibinfo {author} {\bibfnamefont {P.}~\bibnamefont
  {Kehayias}}, \bibinfo {author} {\bibfnamefont {M.~J.}\ \bibnamefont
  {Turner}}, \bibinfo {author} {\bibfnamefont {R.}~\bibnamefont {Trubko}},
  \bibinfo {author} {\bibfnamefont {J.~M.}\ \bibnamefont {Schloss}}, \bibinfo
  {author} {\bibfnamefont {C.~A.}\ \bibnamefont {Hart}}, \bibinfo {author}
  {\bibfnamefont {M.}~\bibnamefont {Wesson}}, \bibinfo {author} {\bibfnamefont
  {D.~R.}\ \bibnamefont {Glenn}},\ and\ \bibinfo {author} {\bibfnamefont
  {R.~L.}\ \bibnamefont {Walsworth}},\ }\bibfield  {title} {\bibinfo {title}
  {Imaging crystal stress in diamond using ensembles of nitrogen-vacancy
  centers},\ }\href {https://doi.org/10.1103/PhysRevB.100.174103} {\bibfield
  {journal} {\bibinfo  {journal} {Phys. Rev. B}\ }\textbf {\bibinfo {volume}
  {100}},\ \bibinfo {pages} {174103} (\bibinfo {year} {2019})}\BibitemShut
  {NoStop}%
\bibitem [{\citenamefont {Kazi}\ \emph {et~al.}(2021)\citenamefont {Kazi},
  \citenamefont {Shelby}, \citenamefont {Watanabe}, \citenamefont {Itoh},
  \citenamefont {Shutthanandan}, \citenamefont {Wiggins},\ and\ \citenamefont
  {Fu}}]{kazi2020}%
  \BibitemOpen
  \bibfield  {author} {\bibinfo {author} {\bibfnamefont {Z.}~\bibnamefont
  {Kazi}}, \bibinfo {author} {\bibfnamefont {I.~M.}\ \bibnamefont {Shelby}},
  \bibinfo {author} {\bibfnamefont {H.}~\bibnamefont {Watanabe}}, \bibinfo
  {author} {\bibfnamefont {K.~M.}\ \bibnamefont {Itoh}}, \bibinfo {author}
  {\bibfnamefont {V.}~\bibnamefont {Shutthanandan}}, \bibinfo {author}
  {\bibfnamefont {P.~A.}\ \bibnamefont {Wiggins}},\ and\ \bibinfo {author}
  {\bibfnamefont {K.-M.~C.}\ \bibnamefont {Fu}},\ }\href@noop {} {\bibinfo
  {title} {Wide-field dynamic magnetic microscopy using double-double quantum
  driving of a diamond defect ensemble}} (\bibinfo {year} {2021}),\ \Eprint
  {https://arxiv.org/abs/2002.06237} {arXiv:2002.06237} \BibitemShut {NoStop}%
\bibitem [{\citenamefont {Fescenko}\ \emph {et~al.}(2020)\citenamefont
  {Fescenko}, \citenamefont {Jarmola}, \citenamefont {Savukov}, \citenamefont
  {Kehayias}, \citenamefont {Smits}, \citenamefont {Damron}, \citenamefont
  {Ristoff}, \citenamefont {Mosavian},\ and\ \citenamefont
  {Acosta}}]{Fescenko2019}%
  \BibitemOpen
  \bibfield  {author} {\bibinfo {author} {\bibfnamefont {I.}~\bibnamefont
  {Fescenko}}, \bibinfo {author} {\bibfnamefont {A.}~\bibnamefont {Jarmola}},
  \bibinfo {author} {\bibfnamefont {I.}~\bibnamefont {Savukov}}, \bibinfo
  {author} {\bibfnamefont {P.}~\bibnamefont {Kehayias}}, \bibinfo {author}
  {\bibfnamefont {J.}~\bibnamefont {Smits}}, \bibinfo {author} {\bibfnamefont
  {J.}~\bibnamefont {Damron}}, \bibinfo {author} {\bibfnamefont
  {N.}~\bibnamefont {Ristoff}}, \bibinfo {author} {\bibfnamefont
  {N.}~\bibnamefont {Mosavian}},\ and\ \bibinfo {author} {\bibfnamefont
  {V.~M.}\ \bibnamefont {Acosta}},\ }\bibfield  {title} {\bibinfo {title}
  {Diamond magnetometer enhanced by ferrite flux concentrators},\ }\href
  {https://doi.org/10.1103/PhysRevResearch.2.023394} {\bibfield  {journal}
  {\bibinfo  {journal} {Phys. Rev. Research}\ }\textbf {\bibinfo {volume}
  {2}},\ \bibinfo {pages} {023394} (\bibinfo {year} {2020})}\BibitemShut
  {NoStop}%
\bibitem [{\citenamefont {Bauch}\ \emph {et~al.}(2018)\citenamefont {Bauch},
  \citenamefont {Hart}, \citenamefont {Schloss}, \citenamefont {Turner},
  \citenamefont {Barry}, \citenamefont {Kehayias}, \citenamefont {Singh},\ and\
  \citenamefont {Walsworth}}]{P1DQ}%
  \BibitemOpen
  \bibfield  {author} {\bibinfo {author} {\bibfnamefont {E.}~\bibnamefont
  {Bauch}}, \bibinfo {author} {\bibfnamefont {C.~A.}\ \bibnamefont {Hart}},
  \bibinfo {author} {\bibfnamefont {J.~M.}\ \bibnamefont {Schloss}}, \bibinfo
  {author} {\bibfnamefont {M.~J.}\ \bibnamefont {Turner}}, \bibinfo {author}
  {\bibfnamefont {J.~F.}\ \bibnamefont {Barry}}, \bibinfo {author}
  {\bibfnamefont {P.}~\bibnamefont {Kehayias}}, \bibinfo {author}
  {\bibfnamefont {S.}~\bibnamefont {Singh}},\ and\ \bibinfo {author}
  {\bibfnamefont {R.~L.}\ \bibnamefont {Walsworth}},\ }\bibfield  {title}
  {\bibinfo {title} {Ultralong dephasing times in solid-state spin ensembles
  via quantum control},\ }\href {https://doi.org/10.1103/PhysRevX.8.031025}
  {\bibfield  {journal} {\bibinfo  {journal} {Phys. Rev. X}\ }\textbf {\bibinfo
  {volume} {8}},\ \bibinfo {pages} {031025} (\bibinfo {year}
  {2018})}\BibitemShut {NoStop}%
\bibitem [{\citenamefont {Mamin}\ \emph {et~al.}(2014)\citenamefont {Mamin},
  \citenamefont {Sherwood}, \citenamefont {Kim}, \citenamefont {Rettner},
  \citenamefont {Ohno}, \citenamefont {Awschalom},\ and\ \citenamefont
  {Rugar}}]{Mamin2014}%
  \BibitemOpen
  \bibfield  {author} {\bibinfo {author} {\bibfnamefont {H.~J.}\ \bibnamefont
  {Mamin}}, \bibinfo {author} {\bibfnamefont {M.~H.}\ \bibnamefont {Sherwood}},
  \bibinfo {author} {\bibfnamefont {M.}~\bibnamefont {Kim}}, \bibinfo {author}
  {\bibfnamefont {C.~T.}\ \bibnamefont {Rettner}}, \bibinfo {author}
  {\bibfnamefont {K.}~\bibnamefont {Ohno}}, \bibinfo {author} {\bibfnamefont
  {D.~D.}\ \bibnamefont {Awschalom}},\ and\ \bibinfo {author} {\bibfnamefont
  {D.}~\bibnamefont {Rugar}},\ }\bibfield  {title} {\bibinfo {title}
  {{Multipulse double-quantum magnetometry with near-surface nitrogen-vacancy
  centers}},\ }\href {https://doi.org/10.1103/PhysRevLett.113.030803}
  {\bibfield  {journal} {\bibinfo  {journal} {Physical Review Letters}\
  }\textbf {\bibinfo {volume} {113}},\ \bibinfo {pages} {030803} (\bibinfo
  {year} {2014})}\BibitemShut {NoStop}%
\bibitem [{\citenamefont {Fang}\ \emph {et~al.}(2013)\citenamefont {Fang},
  \citenamefont {Acosta}, \citenamefont {Santori}, \citenamefont {Huang},
  \citenamefont {Itoh}, \citenamefont {Watanabe}, \citenamefont {Shikata},\
  and\ \citenamefont {Beausoleil}}]{Fang2013}%
  \BibitemOpen
  \bibfield  {author} {\bibinfo {author} {\bibfnamefont {K.}~\bibnamefont
  {Fang}}, \bibinfo {author} {\bibfnamefont {V.~M.}\ \bibnamefont {Acosta}},
  \bibinfo {author} {\bibfnamefont {C.}~\bibnamefont {Santori}}, \bibinfo
  {author} {\bibfnamefont {Z.}~\bibnamefont {Huang}}, \bibinfo {author}
  {\bibfnamefont {K.~M.}\ \bibnamefont {Itoh}}, \bibinfo {author}
  {\bibfnamefont {H.}~\bibnamefont {Watanabe}}, \bibinfo {author}
  {\bibfnamefont {S.}~\bibnamefont {Shikata}},\ and\ \bibinfo {author}
  {\bibfnamefont {R.~G.}\ \bibnamefont {Beausoleil}},\ }\bibfield  {title}
  {\bibinfo {title} {{High-sensitivity magnetometry based on quantum beats in
  diamond nitrogen-vacancy centers}},\ }\href
  {https://doi.org/10.1103/PhysRevLett.110.130802} {\bibfield  {journal}
  {\bibinfo  {journal} {Physical Review Letters}\ }\textbf {\bibinfo {volume}
  {110}},\ \bibinfo {pages} {130802} (\bibinfo {year} {2013})}\BibitemShut
  {NoStop}%
\bibitem [{\citenamefont {Jamonneau}\ \emph {et~al.}(2016)\citenamefont
  {Jamonneau}, \citenamefont {Lesik}, \citenamefont {Tetienne}, \citenamefont
  {Alvizu}, \citenamefont {Mayer}, \citenamefont {Dr{\'{e}}au}, \citenamefont
  {Kosen}, \citenamefont {Roch}, \citenamefont {Pezzagna}, \citenamefont
  {Meijer}, \citenamefont {Teraji}, \citenamefont {Kubo}, \citenamefont
  {Bertet}, \citenamefont {Maze},\ and\ \citenamefont
  {Jacques}}]{Jamonneau2016}%
  \BibitemOpen
  \bibfield  {author} {\bibinfo {author} {\bibfnamefont {P.}~\bibnamefont
  {Jamonneau}}, \bibinfo {author} {\bibfnamefont {M.}~\bibnamefont {Lesik}},
  \bibinfo {author} {\bibfnamefont {J.~P.}\ \bibnamefont {Tetienne}}, \bibinfo
  {author} {\bibfnamefont {I.}~\bibnamefont {Alvizu}}, \bibinfo {author}
  {\bibfnamefont {L.}~\bibnamefont {Mayer}}, \bibinfo {author} {\bibfnamefont
  {A.}~\bibnamefont {Dr{\'{e}}au}}, \bibinfo {author} {\bibfnamefont
  {S.}~\bibnamefont {Kosen}}, \bibinfo {author} {\bibfnamefont {J.-F.}\
  \bibnamefont {Roch}}, \bibinfo {author} {\bibfnamefont {S.}~\bibnamefont
  {Pezzagna}}, \bibinfo {author} {\bibfnamefont {J.}~\bibnamefont {Meijer}},
  \bibinfo {author} {\bibfnamefont {T.}~\bibnamefont {Teraji}}, \bibinfo
  {author} {\bibfnamefont {Y.}~\bibnamefont {Kubo}}, \bibinfo {author}
  {\bibfnamefont {P.}~\bibnamefont {Bertet}}, \bibinfo {author} {\bibfnamefont
  {J.~R.}\ \bibnamefont {Maze}},\ and\ \bibinfo {author} {\bibfnamefont
  {V.}~\bibnamefont {Jacques}},\ }\bibfield  {title} {\bibinfo {title}
  {{Competition between electric field and magnetic field noise in the
  decoherence of a single spin in diamond}},\ }\href
  {https://doi.org/10.1103/PhysRevB.93.024305} {\bibfield  {journal} {\bibinfo
  {journal} {Physical Review B}\ }\textbf {\bibinfo {volume} {93}},\ \bibinfo
  {pages} {024305} (\bibinfo {year} {2016})}\BibitemShut {NoStop}%
\bibitem [{\citenamefont {Barson}\ \emph {et~al.}(2017)\citenamefont {Barson},
  \citenamefont {Peddibhotla}, \citenamefont {Ovartchaiyapong}, \citenamefont
  {Ganesan}, \citenamefont {Taylor}, \citenamefont {Gebert}, \citenamefont
  {Mielens}, \citenamefont {Koslowski}, \citenamefont {Simpson}, \citenamefont
  {McGuinness}, \citenamefont {McCallum}, \citenamefont {Prawer}, \citenamefont
  {Onoda}, \citenamefont {Ohshima}, \citenamefont {Bleszynski~Jayich},
  \citenamefont {Jelezko}, \citenamefont {Manson},\ and\ \citenamefont
  {Doherty}}]{marcusStrainHam}%
  \BibitemOpen
  \bibfield  {author} {\bibinfo {author} {\bibfnamefont {M.~S.~J.}\
  \bibnamefont {Barson}}, \bibinfo {author} {\bibfnamefont {P.}~\bibnamefont
  {Peddibhotla}}, \bibinfo {author} {\bibfnamefont {P.}~\bibnamefont
  {Ovartchaiyapong}}, \bibinfo {author} {\bibfnamefont {K.}~\bibnamefont
  {Ganesan}}, \bibinfo {author} {\bibfnamefont {R.~L.}\ \bibnamefont {Taylor}},
  \bibinfo {author} {\bibfnamefont {M.}~\bibnamefont {Gebert}}, \bibinfo
  {author} {\bibfnamefont {Z.}~\bibnamefont {Mielens}}, \bibinfo {author}
  {\bibfnamefont {B.}~\bibnamefont {Koslowski}}, \bibinfo {author}
  {\bibfnamefont {D.~A.}\ \bibnamefont {Simpson}}, \bibinfo {author}
  {\bibfnamefont {L.~P.}\ \bibnamefont {McGuinness}}, \bibinfo {author}
  {\bibfnamefont {J.}~\bibnamefont {McCallum}}, \bibinfo {author}
  {\bibfnamefont {S.}~\bibnamefont {Prawer}}, \bibinfo {author} {\bibfnamefont
  {S.}~\bibnamefont {Onoda}}, \bibinfo {author} {\bibfnamefont
  {T.}~\bibnamefont {Ohshima}}, \bibinfo {author} {\bibfnamefont {A.~C.}\
  \bibnamefont {Bleszynski~Jayich}}, \bibinfo {author} {\bibfnamefont
  {F.}~\bibnamefont {Jelezko}}, \bibinfo {author} {\bibfnamefont {N.~B.}\
  \bibnamefont {Manson}},\ and\ \bibinfo {author} {\bibfnamefont {M.~W.}\
  \bibnamefont {Doherty}},\ }\bibfield  {title} {\bibinfo {title}
  {Nanomechanical sensing using spins in diamond},\ }\href
  {https://doi.org/10.1021/acs.nanolett.6b04544} {\bibfield  {journal}
  {\bibinfo  {journal} {Nano Letters}\ }\textbf {\bibinfo {volume} {17}},\
  \bibinfo {pages} {1496} (\bibinfo {year} {2017})}\BibitemShut {NoStop}%
\bibitem [{\citenamefont {Udvarhelyi}\ \emph {et~al.}(2018)\citenamefont
  {Udvarhelyi}, \citenamefont {Shkolnikov}, \citenamefont {Gali}, \citenamefont
  {Burkard},\ and\ \citenamefont {P\'alyi}}]{galiSpinStrain}%
  \BibitemOpen
  \bibfield  {author} {\bibinfo {author} {\bibfnamefont {P.}~\bibnamefont
  {Udvarhelyi}}, \bibinfo {author} {\bibfnamefont {V.~O.}\ \bibnamefont
  {Shkolnikov}}, \bibinfo {author} {\bibfnamefont {A.}~\bibnamefont {Gali}},
  \bibinfo {author} {\bibfnamefont {G.}~\bibnamefont {Burkard}},\ and\ \bibinfo
  {author} {\bibfnamefont {A.}~\bibnamefont {P\'alyi}},\ }\bibfield  {title}
  {\bibinfo {title} {Spin-strain interaction in nitrogen-vacancy centers in
  diamond},\ }\href {https://doi.org/10.1103/PhysRevB.98.075201} {\bibfield
  {journal} {\bibinfo  {journal} {Phys. Rev. B}\ }\textbf {\bibinfo {volume}
  {98}},\ \bibinfo {pages} {075201} (\bibinfo {year} {2018})}\BibitemShut
  {NoStop}%
\bibitem [{\citenamefont {Barfuss}\ \emph {et~al.}(2019)\citenamefont
  {Barfuss}, \citenamefont {Kasperczyk}, \citenamefont {K\"olbl},\ and\
  \citenamefont {Maletinsky}}]{maletinskyStrainTerms}%
  \BibitemOpen
  \bibfield  {author} {\bibinfo {author} {\bibfnamefont {A.}~\bibnamefont
  {Barfuss}}, \bibinfo {author} {\bibfnamefont {M.}~\bibnamefont {Kasperczyk}},
  \bibinfo {author} {\bibfnamefont {J.}~\bibnamefont {K\"olbl}},\ and\ \bibinfo
  {author} {\bibfnamefont {P.}~\bibnamefont {Maletinsky}},\ }\bibfield  {title}
  {\bibinfo {title} {Spin-stress and spin-strain coupling in diamond-based
  hybrid spin oscillator systems},\ }\href
  {https://doi.org/10.1103/PhysRevB.99.174102} {\bibfield  {journal} {\bibinfo
  {journal} {Phys. Rev. B}\ }\textbf {\bibinfo {volume} {99}},\ \bibinfo
  {pages} {174102} (\bibinfo {year} {2019})}\BibitemShut {NoStop}%
\bibitem [{\citenamefont {Dolde}\ \emph {et~al.}(2011)\citenamefont {Dolde},
  \citenamefont {Fedder}, \citenamefont {Doherty}, \citenamefont
  {N{\"{o}}bauer}, \citenamefont {Rempp}, \citenamefont {Balasubramanian},
  \citenamefont {Wolf}, \citenamefont {Reinhard}, \citenamefont {Hollenberg},
  \citenamefont {Jelezko},\ and\ \citenamefont {Wrachtrup}}]{Dolde2011}%
  \BibitemOpen
  \bibfield  {author} {\bibinfo {author} {\bibfnamefont {F.}~\bibnamefont
  {Dolde}}, \bibinfo {author} {\bibfnamefont {H.}~\bibnamefont {Fedder}},
  \bibinfo {author} {\bibfnamefont {M.~W.}\ \bibnamefont {Doherty}}, \bibinfo
  {author} {\bibfnamefont {T.}~\bibnamefont {N{\"{o}}bauer}}, \bibinfo {author}
  {\bibfnamefont {F.}~\bibnamefont {Rempp}}, \bibinfo {author} {\bibfnamefont
  {G.}~\bibnamefont {Balasubramanian}}, \bibinfo {author} {\bibfnamefont
  {T.}~\bibnamefont {Wolf}}, \bibinfo {author} {\bibfnamefont {F.}~\bibnamefont
  {Reinhard}}, \bibinfo {author} {\bibfnamefont {L.~C.~L.}\ \bibnamefont
  {Hollenberg}}, \bibinfo {author} {\bibfnamefont {F.}~\bibnamefont
  {Jelezko}},\ and\ \bibinfo {author} {\bibfnamefont {J.}~\bibnamefont
  {Wrachtrup}},\ }\bibfield  {title} {\bibinfo {title} {{Electric-field sensing
  using single diamond spins}},\ }\href {https://doi.org/10.1038/nphys1969}
  {\bibfield  {journal} {\bibinfo  {journal} {Nature Physics}\ }\textbf
  {\bibinfo {volume} {7}},\ \bibinfo {pages} {459} (\bibinfo {year}
  {2011})}\BibitemShut {NoStop}%
\bibitem [{sup()}]{suppl}%
  \BibitemOpen
  \href@noop {} {}\bibinfo {note} {See the Supplemental Material at [URL to be
  inserted] for additional information on the NV Hamiltonian, DQ 4-Ramsey
  measurement protocol, and magnetic imaging results.}\BibitemShut {Stop}%
\bibitem [{\citenamefont {Friel}\ \emph {et~al.}(2009)\citenamefont {Friel},
  \citenamefont {Clewes}, \citenamefont {Dhillon}, \citenamefont {Perkins},
  \citenamefont {Twitchen},\ and\ \citenamefont {Scarsbrook}}]{Friel2009}%
  \BibitemOpen
  \bibfield  {author} {\bibinfo {author} {\bibfnamefont {I.}~\bibnamefont
  {Friel}}, \bibinfo {author} {\bibfnamefont {S.~L.}\ \bibnamefont {Clewes}},
  \bibinfo {author} {\bibfnamefont {H.~K.}\ \bibnamefont {Dhillon}}, \bibinfo
  {author} {\bibfnamefont {N.}~\bibnamefont {Perkins}}, \bibinfo {author}
  {\bibfnamefont {D.~J.}\ \bibnamefont {Twitchen}},\ and\ \bibinfo {author}
  {\bibfnamefont {G.~A.}\ \bibnamefont {Scarsbrook}},\ }\bibfield  {title}
  {\bibinfo {title} {{Control of surface and bulk crystalline quality in single
  crystal diamond grown by chemical vapour deposition}},\ }\href
  {https://doi.org/10.1016/j.diamond.2009.01.013} {\bibfield  {journal}
  {\bibinfo  {journal} {Diamond and Related Materials}\ }\textbf {\bibinfo
  {volume} {18}},\ \bibinfo {pages} {808} (\bibinfo {year} {2009})}\BibitemShut
  {NoStop}%
\bibitem [{\citenamefont {Bar-Gill}\ \emph {et~al.}(2013)\citenamefont
  {Bar-Gill}, \citenamefont {Pham}, \citenamefont {Jarmola}, \citenamefont
  {Budker},\ and\ \citenamefont {Walsworth}}]{Bar-Gill2013}%
  \BibitemOpen
  \bibfield  {author} {\bibinfo {author} {\bibfnamefont {N.}~\bibnamefont
  {Bar-Gill}}, \bibinfo {author} {\bibfnamefont {L.~M.}\ \bibnamefont {Pham}},
  \bibinfo {author} {\bibfnamefont {A.}~\bibnamefont {Jarmola}}, \bibinfo
  {author} {\bibfnamefont {D.}~\bibnamefont {Budker}},\ and\ \bibinfo {author}
  {\bibfnamefont {R.~L.}\ \bibnamefont {Walsworth}},\ }\bibfield  {title}
  {\bibinfo {title} {{Solid-state electronic spin coherence time approaching
  one second.}},\ }\href {https://doi.org/10.1038/ncomms2771} {\bibfield
  {journal} {\bibinfo  {journal} {Nature Communications}\ }\textbf {\bibinfo
  {volume} {4}},\ \bibinfo {pages} {1743} (\bibinfo {year} {2013})}\BibitemShut
  {NoStop}%
\bibitem [{\citenamefont {Przybylski}\ \emph {et~al.}(2017)\citenamefont
  {Przybylski}, \citenamefont {Thiel}, \citenamefont {Keller-Findeisen},
  \citenamefont {Stock},\ and\ \citenamefont {Bates}}]{GPUfit}%
  \BibitemOpen
  \bibfield  {author} {\bibinfo {author} {\bibfnamefont {A.}~\bibnamefont
  {Przybylski}}, \bibinfo {author} {\bibfnamefont {B.}~\bibnamefont {Thiel}},
  \bibinfo {author} {\bibfnamefont {J.}~\bibnamefont {Keller-Findeisen}},
  \bibinfo {author} {\bibfnamefont {B.}~\bibnamefont {Stock}},\ and\ \bibinfo
  {author} {\bibfnamefont {M.}~\bibnamefont {Bates}},\ }\bibfield  {title}
  {\bibinfo {title} {{Gpufit: An open-source toolkit for GPU-accelerated curve
  fitting}},\ }\href {https://doi.org/10.1038/s41598-017-15313-9} {\bibfield
  {journal} {\bibinfo  {journal} {Scientific Reports}\ }\textbf {\bibinfo
  {volume} {7}},\ \bibinfo {pages} {15722} (\bibinfo {year}
  {2017})}\BibitemShut {NoStop}%
\bibitem [{\citenamefont {Taylor}\ \emph {et~al.}(2008)\citenamefont {Taylor},
  \citenamefont {Cappellaro}, \citenamefont {Childress}, \citenamefont {Jiang},
  \citenamefont {Budker}, \citenamefont {Hemmer}, \citenamefont {Yacoby},
  \citenamefont {Walsworth},\ and\ \citenamefont {Lukin}}]{Taylor2008}%
  \BibitemOpen
  \bibfield  {author} {\bibinfo {author} {\bibfnamefont {J.~M.}\ \bibnamefont
  {Taylor}}, \bibinfo {author} {\bibfnamefont {P.}~\bibnamefont {Cappellaro}},
  \bibinfo {author} {\bibfnamefont {L.}~\bibnamefont {Childress}}, \bibinfo
  {author} {\bibfnamefont {L.}~\bibnamefont {Jiang}}, \bibinfo {author}
  {\bibfnamefont {D.}~\bibnamefont {Budker}}, \bibinfo {author} {\bibfnamefont
  {P.~R.}\ \bibnamefont {Hemmer}}, \bibinfo {author} {\bibfnamefont
  {A.}~\bibnamefont {Yacoby}}, \bibinfo {author} {\bibfnamefont
  {R.}~\bibnamefont {Walsworth}},\ and\ \bibinfo {author} {\bibfnamefont
  {M.~D.}\ \bibnamefont {Lukin}},\ }\bibfield  {title} {\bibinfo {title}
  {{High-sensitivity diamond magnetometer with nanoscale resolution}},\ }\href
  {https://doi.org/10.1038/nphys1075} {\bibfield  {journal} {\bibinfo
  {journal} {Nature Physics}\ }\textbf {\bibinfo {volume} {4}},\ \bibinfo
  {pages} {810} (\bibinfo {year} {2008})}\BibitemShut {NoStop}%
\bibitem [{\citenamefont {{Le Sage}}\ \emph {et~al.}(2012)\citenamefont {{Le
  Sage}}, \citenamefont {Pham}, \citenamefont {Bar-Gill}, \citenamefont
  {Belthangady}, \citenamefont {Lukin}, \citenamefont {Yacoby},\ and\
  \citenamefont {Walsworth}}]{LeSage2012}%
  \BibitemOpen
  \bibfield  {author} {\bibinfo {author} {\bibfnamefont {D.}~\bibnamefont {{Le
  Sage}}}, \bibinfo {author} {\bibfnamefont {L.~M.}\ \bibnamefont {Pham}},
  \bibinfo {author} {\bibfnamefont {N.}~\bibnamefont {Bar-Gill}}, \bibinfo
  {author} {\bibfnamefont {C.}~\bibnamefont {Belthangady}}, \bibinfo {author}
  {\bibfnamefont {M.~D.}\ \bibnamefont {Lukin}}, \bibinfo {author}
  {\bibfnamefont {A.}~\bibnamefont {Yacoby}},\ and\ \bibinfo {author}
  {\bibfnamefont {R.~L.}\ \bibnamefont {Walsworth}},\ }\bibfield  {title}
  {\bibinfo {title} {{Efficient photon detection from color centers in a
  diamond optical waveguide}},\ }\href
  {https://doi.org/10.1103/PhysRevB.85.121202} {\bibfield  {journal} {\bibinfo
  {journal} {Physical Review B}\ }\textbf {\bibinfo {volume} {85}},\ \bibinfo
  {pages} {121202(R)} (\bibinfo {year} {2012})}\BibitemShut {NoStop}%
\bibitem [{\citenamefont {Bal}\ \emph {et~al.}(2012)\citenamefont {Bal},
  \citenamefont {Deng}, \citenamefont {Orgiazzi}, \citenamefont {Ong},\ and\
  \citenamefont {Lupascu}}]{Bal2012}%
  \BibitemOpen
  \bibfield  {author} {\bibinfo {author} {\bibfnamefont {M.}~\bibnamefont
  {Bal}}, \bibinfo {author} {\bibfnamefont {C.}~\bibnamefont {Deng}}, \bibinfo
  {author} {\bibfnamefont {J.~L.}\ \bibnamefont {Orgiazzi}}, \bibinfo {author}
  {\bibfnamefont {F.~R.}\ \bibnamefont {Ong}},\ and\ \bibinfo {author}
  {\bibfnamefont {A.}~\bibnamefont {Lupascu}},\ }\bibfield  {title} {\bibinfo
  {title} {{Ultrasensitive magnetic field detection using a single artificial
  atom}},\ }\href {https://doi.org/10.1038/ncomms2332} {\bibfield  {journal}
  {\bibinfo  {journal} {Nature Communications}\ }\textbf {\bibinfo {volume}
  {3}},\ \bibinfo {pages} {1324} (\bibinfo {year} {2012})}\BibitemShut
  {NoStop}%
\bibitem [{\citenamefont {Schoenfeld}\ and\ \citenamefont
  {Harneit}(2011)}]{Schoenfeld2011}%
  \BibitemOpen
  \bibfield  {author} {\bibinfo {author} {\bibfnamefont {R.~S.}\ \bibnamefont
  {Schoenfeld}}\ and\ \bibinfo {author} {\bibfnamefont {W.}~\bibnamefont
  {Harneit}},\ }\bibfield  {title} {\bibinfo {title} {{Real time magnetic field
  sensing and imaging using a single spin in diamond}},\ }\href
  {https://doi.org/10.1103/PhysRevLett.106.030802} {\bibfield  {journal}
  {\bibinfo  {journal} {Physical Review Letters}\ }\textbf {\bibinfo {volume}
  {106}},\ \bibinfo {pages} {030802} (\bibinfo {year} {2011})}\BibitemShut
  {NoStop}%
\bibitem [{\citenamefont {Schloss}\ \emph {et~al.}(2018)\citenamefont
  {Schloss}, \citenamefont {Barry}, \citenamefont {Turner},\ and\ \citenamefont
  {Walsworth}}]{SVMPaper}%
  \BibitemOpen
  \bibfield  {author} {\bibinfo {author} {\bibfnamefont {J.~M.}\ \bibnamefont
  {Schloss}}, \bibinfo {author} {\bibfnamefont {J.~F.}\ \bibnamefont {Barry}},
  \bibinfo {author} {\bibfnamefont {M.~J.}\ \bibnamefont {Turner}},\ and\
  \bibinfo {author} {\bibfnamefont {R.~L.}\ \bibnamefont {Walsworth}},\
  }\bibfield  {title} {\bibinfo {title} {{Simultaneous Broadband Vector
  Magnetometry Using Solid-State Spins}},\ }\href
  {https://doi.org/10.1103/PhysRevApplied.10.034044} {\bibfield  {journal}
  {\bibinfo  {journal} {Physical Review Applied}\ }\textbf {\bibinfo {volume}
  {10}},\ \bibinfo {pages} {034044} (\bibinfo {year} {2018})}\BibitemShut
  {NoStop}%
\bibitem [{\citenamefont {Barry}\ \emph {et~al.}(2016)\citenamefont {Barry},
  \citenamefont {Turner}, \citenamefont {Schloss}, \citenamefont {Glenn},
  \citenamefont {Song}, \citenamefont {Lukin}, \citenamefont {Park},\ and\
  \citenamefont {Walsworth}}]{NeuronPaper}%
  \BibitemOpen
  \bibfield  {author} {\bibinfo {author} {\bibfnamefont {J.~F.}\ \bibnamefont
  {Barry}}, \bibinfo {author} {\bibfnamefont {M.~J.}\ \bibnamefont {Turner}},
  \bibinfo {author} {\bibfnamefont {J.~M.}\ \bibnamefont {Schloss}}, \bibinfo
  {author} {\bibfnamefont {D.~R.}\ \bibnamefont {Glenn}}, \bibinfo {author}
  {\bibfnamefont {Y.}~\bibnamefont {Song}}, \bibinfo {author} {\bibfnamefont
  {M.~D.}\ \bibnamefont {Lukin}}, \bibinfo {author} {\bibfnamefont
  {H.}~\bibnamefont {Park}},\ and\ \bibinfo {author} {\bibfnamefont {R.~L.}\
  \bibnamefont {Walsworth}},\ }\bibfield  {title} {\bibinfo {title} {{Optical
  magnetic detection of single-neuron action potentials using quantum defects
  in diamond}},\ }\href {https://doi.org/10.1073/pnas.1601513113} {\bibfield
  {journal} {\bibinfo  {journal} {Proceedings of the National Academy of
  Sciences of the United States of America}\ }\textbf {\bibinfo {volume}
  {113}},\ \bibinfo {pages} {14133} (\bibinfo {year} {2016})}\BibitemShut
  {NoStop}%
\bibitem [{\citenamefont {Canciani}\ and\ \citenamefont
  {Raquet}(2016)}]{canciani_absolute_2016}%
  \BibitemOpen
  \bibfield  {author} {\bibinfo {author} {\bibfnamefont {A.}~\bibnamefont
  {Canciani}}\ and\ \bibinfo {author} {\bibfnamefont {J.}~\bibnamefont
  {Raquet}},\ }\bibfield  {title} {\bibinfo {title} {{Absolute Positioning
  Using the Earth's Magnetic Anomaly Field}},\ }\href
  {https://doi.org/10.1002/navi.138} {\bibfield  {journal} {\bibinfo  {journal}
  {Navigation, Journal of the Institute of Navigation}\ }\textbf {\bibinfo
  {volume} {63}},\ \bibinfo {pages} {111} (\bibinfo {year} {2016})}\BibitemShut
  {NoStop}%
\bibitem [{\citenamefont {Wee}\ \emph {et~al.}(2007)\citenamefont {Wee},
  \citenamefont {Tzeng}, \citenamefont {Han}, \citenamefont {Chang},
  \citenamefont {Fann}, \citenamefont {Hsu}, \citenamefont {Chen},\ and\
  \citenamefont {Yu}}]{Wee2007}%
  \BibitemOpen
  \bibfield  {author} {\bibinfo {author} {\bibfnamefont {T.-L.}\ \bibnamefont
  {Wee}}, \bibinfo {author} {\bibfnamefont {Y.-K.}\ \bibnamefont {Tzeng}},
  \bibinfo {author} {\bibfnamefont {C.-C.}\ \bibnamefont {Han}}, \bibinfo
  {author} {\bibfnamefont {H.-C.}\ \bibnamefont {Chang}}, \bibinfo {author}
  {\bibfnamefont {W.}~\bibnamefont {Fann}}, \bibinfo {author} {\bibfnamefont
  {J.-H.}\ \bibnamefont {Hsu}}, \bibinfo {author} {\bibfnamefont {K.-M.}\
  \bibnamefont {Chen}},\ and\ \bibinfo {author} {\bibfnamefont {Y.-C.}\
  \bibnamefont {Yu}},\ }\bibfield  {title} {\bibinfo {title} {{Two-photon
  excited fluorescence of nitrogen-vacancy centers in proton-irradiated type Ib
  diamond}},\ }\href {https://doi.org/10.1021/jp073938o} {\bibfield  {journal}
  {\bibinfo  {journal} {Journal of Physical Chemistry A}\ }\textbf {\bibinfo
  {volume} {111}},\ \bibinfo {pages} {9379} (\bibinfo {year}
  {2007})}\BibitemShut {NoStop}%
\bibitem [{\citenamefont {Alsid}\ \emph {et~al.}(2019)\citenamefont {Alsid},
  \citenamefont {Barry}, \citenamefont {Pham}, \citenamefont {Schloss},
  \citenamefont {O'Keeffe}, \citenamefont {Cappellaro},\ and\ \citenamefont
  {Braje}}]{alsid_photoluminescence_2019}%
  \BibitemOpen
  \bibfield  {author} {\bibinfo {author} {\bibfnamefont {S.~T.}\ \bibnamefont
  {Alsid}}, \bibinfo {author} {\bibfnamefont {J.~F.}\ \bibnamefont {Barry}},
  \bibinfo {author} {\bibfnamefont {L.~M.}\ \bibnamefont {Pham}}, \bibinfo
  {author} {\bibfnamefont {J.~M.}\ \bibnamefont {Schloss}}, \bibinfo {author}
  {\bibfnamefont {M.~F.}\ \bibnamefont {O'Keeffe}}, \bibinfo {author}
  {\bibfnamefont {P.}~\bibnamefont {Cappellaro}},\ and\ \bibinfo {author}
  {\bibfnamefont {D.~A.}\ \bibnamefont {Braje}},\ }\bibfield  {title} {\bibinfo
  {title} {{Photoluminescence Decomposition Analysis: A Technique to
  Characterize N - V Creation in Diamond}},\ }\href
  {https://doi.org/10.1103/PhysRevApplied.12.044003} {\bibfield  {journal}
  {\bibinfo  {journal} {Physical Review Applied}\ }\textbf {\bibinfo {volume}
  {12}},\ \bibinfo {pages} {044003} (\bibinfo {year} {2019})}\BibitemShut
  {NoStop}%
\bibitem [{\citenamefont {Aude~Craik}\ \emph {et~al.}(2020)\citenamefont
  {Aude~Craik}, \citenamefont {Kehayias}, \citenamefont {Greenspon},
  \citenamefont {Zhang}, \citenamefont {Turner}, \citenamefont {Schloss},
  \citenamefont {Bauch}, \citenamefont {Hart}, \citenamefont {Hu},\ and\
  \citenamefont {Walsworth}}]{aude_craik_microwave-assisted_2018}%
  \BibitemOpen
  \bibfield  {author} {\bibinfo {author} {\bibfnamefont {D.}~\bibnamefont
  {Aude~Craik}}, \bibinfo {author} {\bibfnamefont {P.}~\bibnamefont
  {Kehayias}}, \bibinfo {author} {\bibfnamefont {A.}~\bibnamefont {Greenspon}},
  \bibinfo {author} {\bibfnamefont {X.}~\bibnamefont {Zhang}}, \bibinfo
  {author} {\bibfnamefont {M.}~\bibnamefont {Turner}}, \bibinfo {author}
  {\bibfnamefont {J.}~\bibnamefont {Schloss}}, \bibinfo {author} {\bibfnamefont
  {E.}~\bibnamefont {Bauch}}, \bibinfo {author} {\bibfnamefont
  {C.}~\bibnamefont {Hart}}, \bibinfo {author} {\bibfnamefont {E.}~\bibnamefont
  {Hu}},\ and\ \bibinfo {author} {\bibfnamefont {R.}~\bibnamefont
  {Walsworth}},\ }\bibfield  {title} {\bibinfo {title} {{Microwave-Assisted
  Spectroscopy Technique for Studying Charge State in Nitrogen-Vacancy
  Ensembles in Diamond}},\ }\href
  {https://doi.org/10.1103/PhysRevApplied.14.014009} {\bibfield  {journal}
  {\bibinfo  {journal} {Physical Review Applied}\ }\textbf {\bibinfo {volume}
  {14}},\ \bibinfo {pages} {014009} (\bibinfo {year} {2020})}\BibitemShut
  {NoStop}%
\bibitem [{\citenamefont {Edmonds}\ \emph {et~al.}(2021)\citenamefont
  {Edmonds}, \citenamefont {Hart}, \citenamefont {Turner}, \citenamefont
  {Colard}, \citenamefont {Schloss}, \citenamefont {Olsson}, \citenamefont
  {Trubko}, \citenamefont {Markham}, \citenamefont {Rathmill}, \citenamefont
  {Horne-Smith}, \citenamefont {Lew}, \citenamefont {Manickam}, \citenamefont
  {Bruce}, \citenamefont {Kaup}, \citenamefont {Russo}, \citenamefont
  {DiMario}, \citenamefont {South}, \citenamefont {Hansen}, \citenamefont
  {Twitchen},\ and\ \citenamefont {Walsworth}}]{purpleDiamond}%
  \BibitemOpen
  \bibfield  {author} {\bibinfo {author} {\bibfnamefont {A.~M.}\ \bibnamefont
  {Edmonds}}, \bibinfo {author} {\bibfnamefont {C.~A.}\ \bibnamefont {Hart}},
  \bibinfo {author} {\bibfnamefont {M.~J.}\ \bibnamefont {Turner}}, \bibinfo
  {author} {\bibfnamefont {P.-O.}\ \bibnamefont {Colard}}, \bibinfo {author}
  {\bibfnamefont {J.~M.}\ \bibnamefont {Schloss}}, \bibinfo {author}
  {\bibfnamefont {K.}~\bibnamefont {Olsson}}, \bibinfo {author} {\bibfnamefont
  {R.}~\bibnamefont {Trubko}}, \bibinfo {author} {\bibfnamefont {M.~L.}\
  \bibnamefont {Markham}}, \bibinfo {author} {\bibfnamefont {A.}~\bibnamefont
  {Rathmill}}, \bibinfo {author} {\bibfnamefont {B.}~\bibnamefont
  {Horne-Smith}}, \bibinfo {author} {\bibfnamefont {W.}~\bibnamefont {Lew}},
  \bibinfo {author} {\bibfnamefont {A.}~\bibnamefont {Manickam}}, \bibinfo
  {author} {\bibfnamefont {S.}~\bibnamefont {Bruce}}, \bibinfo {author}
  {\bibfnamefont {P.~G.}\ \bibnamefont {Kaup}}, \bibinfo {author}
  {\bibfnamefont {J.~C.}\ \bibnamefont {Russo}}, \bibinfo {author}
  {\bibfnamefont {M.~J.}\ \bibnamefont {DiMario}}, \bibinfo {author}
  {\bibfnamefont {J.~T.}\ \bibnamefont {South}}, \bibinfo {author}
  {\bibfnamefont {J.~T.}\ \bibnamefont {Hansen}}, \bibinfo {author}
  {\bibfnamefont {D.~J.}\ \bibnamefont {Twitchen}},\ and\ \bibinfo {author}
  {\bibfnamefont {R.}~\bibnamefont {Walsworth}},\ }\bibfield  {title} {\bibinfo
  {title} {Characterisation of {CVD} diamond with high concentrations of
  nitrogen for magnetic-field sensing applications},\ }\bibfield  {journal}
  {\bibinfo  {journal} {Materials for Quantum Technology}\ }\href
  {https://doi.org/10.1088/2633-4356/abd88a} {10.1088/2633-4356/abd88a}
  (\bibinfo {year} {2021})\BibitemShut {NoStop}%
\bibitem [{\citenamefont {Toyli}\ \emph {et~al.}(2013)\citenamefont {Toyli},
  \citenamefont {de~las Casas}, \citenamefont {Christle}, \citenamefont
  {Dobrovitski},\ and\ \citenamefont {Awschalom}}]{Toyli2013}%
  \BibitemOpen
  \bibfield  {author} {\bibinfo {author} {\bibfnamefont {D.~M.}\ \bibnamefont
  {Toyli}}, \bibinfo {author} {\bibfnamefont {C.~F.}\ \bibnamefont {de~las
  Casas}}, \bibinfo {author} {\bibfnamefont {D.~J.}\ \bibnamefont {Christle}},
  \bibinfo {author} {\bibfnamefont {V.~V.}\ \bibnamefont {Dobrovitski}},\ and\
  \bibinfo {author} {\bibfnamefont {D.~D.}\ \bibnamefont {Awschalom}},\
  }\bibfield  {title} {\bibinfo {title} {Fluorescence thermometry enhanced by
  the quantum coherence of single spins in diamond},\ }\href
  {https://doi.org/10.1073/pnas.1306825110} {\bibfield  {journal} {\bibinfo
  {journal} {Proceedings of the National Academy of Sciences}\ }\textbf
  {\bibinfo {volume} {110}},\ \bibinfo {pages} {8417} (\bibinfo {year}
  {2013})}\BibitemShut {NoStop}%
\bibitem [{\citenamefont {Hodges}\ \emph {et~al.}(2013)\citenamefont {Hodges},
  \citenamefont {Yao}, \citenamefont {Maclaurin}, \citenamefont {Rastogi},
  \citenamefont {Lukin},\ and\ \citenamefont {Englund}}]{Hodges}%
  \BibitemOpen
  \bibfield  {author} {\bibinfo {author} {\bibfnamefont {J.~S.}\ \bibnamefont
  {Hodges}}, \bibinfo {author} {\bibfnamefont {N.~Y.}\ \bibnamefont {Yao}},
  \bibinfo {author} {\bibfnamefont {D.}~\bibnamefont {Maclaurin}}, \bibinfo
  {author} {\bibfnamefont {C.}~\bibnamefont {Rastogi}}, \bibinfo {author}
  {\bibfnamefont {M.~D.}\ \bibnamefont {Lukin}},\ and\ \bibinfo {author}
  {\bibfnamefont {D.}~\bibnamefont {Englund}},\ }\bibfield  {title} {\bibinfo
  {title} {{Timekeeping with electron spin states in diamond}},\ }\href
  {https://doi.org/10.1103/PhysRevA.87.032118} {\bibfield  {journal} {\bibinfo
  {journal} {Physical Review A}\ }\textbf {\bibinfo {volume} {87}},\ \bibinfo
  {pages} {032118} (\bibinfo {year} {2013})}\BibitemShut {NoStop}%
\bibitem [{\citenamefont {Rajendran}\ \emph {et~al.}(2017)\citenamefont
  {Rajendran}, \citenamefont {Zobrist}, \citenamefont {Sushkov}, \citenamefont
  {Walsworth},\ and\ \citenamefont {Lukin}}]{Rajendran2017a}%
  \BibitemOpen
  \bibfield  {author} {\bibinfo {author} {\bibfnamefont {S.}~\bibnamefont
  {Rajendran}}, \bibinfo {author} {\bibfnamefont {N.}~\bibnamefont {Zobrist}},
  \bibinfo {author} {\bibfnamefont {A.~O.}\ \bibnamefont {Sushkov}}, \bibinfo
  {author} {\bibfnamefont {R.}~\bibnamefont {Walsworth}},\ and\ \bibinfo
  {author} {\bibfnamefont {M.}~\bibnamefont {Lukin}},\ }\bibfield  {title}
  {\bibinfo {title} {{A method for directional detection of dark matter using
  spectroscopy of crystal defects}},\ }\href
  {https://doi.org/10.1103/PhysRevD.96.035009} {\bibfield  {journal} {\bibinfo
  {journal} {Physical Review D}\ }\textbf {\bibinfo {volume} {96}},\ \bibinfo
  {pages} {035009} (\bibinfo {year} {2017})}\BibitemShut {NoStop}%
\end{thebibliography}%


\begin{thebibliography}{7}%
\makeatletter
\providecommand \@ifxundefined [1]{%
 \@ifx{#1\undefined}
}%
\providecommand \@ifnum [1]{%
 \ifnum #1\expandafter \@firstoftwo
 \else \expandafter \@secondoftwo
 \fi
}%
\providecommand \@ifx [1]{%
 \ifx #1\expandafter \@firstoftwo
 \else \expandafter \@secondoftwo
 \fi
}%
\providecommand \natexlab [1]{#1}%
\providecommand \enquote  [1]{``#1''}%
\providecommand \bibnamefont  [1]{#1}%
\providecommand \bibfnamefont [1]{#1}%
\providecommand \citenamefont [1]{#1}%
\providecommand \href@noop [0]{\@secondoftwo}%
\providecommand \href [0]{\begingroup \@sanitize@url \@href}%
\providecommand \@href[1]{\@@startlink{#1}\@@href}%
\providecommand \@@href[1]{\endgroup#1\@@endlink}%
\providecommand \@sanitize@url [0]{\catcode `\\12\catcode `\$12\catcode
  `\&12\catcode `\#12\catcode `\^12\catcode `\_12\catcode `\%12\relax}%
\providecommand \@@startlink[1]{}%
\providecommand \@@endlink[0]{}%
\providecommand \url  [0]{\begingroup\@sanitize@url \@url }%
\providecommand \@url [1]{\endgroup\@href {#1}{\urlprefix }}%
\providecommand \urlprefix  [0]{URL }%
\providecommand \Eprint [0]{\href }%
\providecommand \doibase [0]{https://doi.org/}%
\providecommand \selectlanguage [0]{\@gobble}%
\providecommand \bibinfo  [0]{\@secondoftwo}%
\providecommand \bibfield  [0]{\@secondoftwo}%
\providecommand \translation [1]{[#1]}%
\providecommand \BibitemOpen [0]{}%
\providecommand \bibitemStop [0]{}%
\providecommand \bibitemNoStop [0]{.\EOS\space}%
\providecommand \EOS [0]{\spacefactor3000\relax}%
\providecommand \BibitemShut  [1]{\csname bibitem#1\endcsname}%
\let\auto@bib@innerbib\@empty
\bibitem [{\citenamefont {Barson}\ \emph {et~al.}(2017)\citenamefont {Barson},
  \citenamefont {Peddibhotla}, \citenamefont {Ovartchaiyapong}, \citenamefont
  {Ganesan}, \citenamefont {Taylor}, \citenamefont {Gebert}, \citenamefont
  {Mielens}, \citenamefont {Koslowski}, \citenamefont {Simpson}, \citenamefont
  {McGuinness}, \citenamefont {McCallum}, \citenamefont {Prawer}, \citenamefont
  {Onoda}, \citenamefont {Ohshima}, \citenamefont {Bleszynski~Jayich},
  \citenamefont {Jelezko}, \citenamefont {Manson},\ and\ \citenamefont
  {Doherty}}]{marcusStrainHam}%
  \BibitemOpen
  \bibfield  {author} {\bibinfo {author} {\bibfnamefont {M.~S.~J.}\
  \bibnamefont {Barson}}, \bibinfo {author} {\bibfnamefont {P.}~\bibnamefont
  {Peddibhotla}}, \bibinfo {author} {\bibfnamefont {P.}~\bibnamefont
  {Ovartchaiyapong}}, \bibinfo {author} {\bibfnamefont {K.}~\bibnamefont
  {Ganesan}}, \bibinfo {author} {\bibfnamefont {R.~L.}\ \bibnamefont {Taylor}},
  \bibinfo {author} {\bibfnamefont {M.}~\bibnamefont {Gebert}}, \bibinfo
  {author} {\bibfnamefont {Z.}~\bibnamefont {Mielens}}, \bibinfo {author}
  {\bibfnamefont {B.}~\bibnamefont {Koslowski}}, \bibinfo {author}
  {\bibfnamefont {D.~A.}\ \bibnamefont {Simpson}}, \bibinfo {author}
  {\bibfnamefont {L.~P.}\ \bibnamefont {McGuinness}}, \bibinfo {author}
  {\bibfnamefont {J.}~\bibnamefont {McCallum}}, \bibinfo {author}
  {\bibfnamefont {S.}~\bibnamefont {Prawer}}, \bibinfo {author} {\bibfnamefont
  {S.}~\bibnamefont {Onoda}}, \bibinfo {author} {\bibfnamefont
  {T.}~\bibnamefont {Ohshima}}, \bibinfo {author} {\bibfnamefont {A.~C.}\
  \bibnamefont {Bleszynski~Jayich}}, \bibinfo {author} {\bibfnamefont
  {F.}~\bibnamefont {Jelezko}}, \bibinfo {author} {\bibfnamefont {N.~B.}\
  \bibnamefont {Manson}},\ and\ \bibinfo {author} {\bibfnamefont {M.~W.}\
  \bibnamefont {Doherty}},\ }\bibfield  {title} {\bibinfo {title}
  {Nanomechanical sensing using spins in diamond},\ }\href
  {https://doi.org/10.1021/acs.nanolett.6b04544} {\bibfield  {journal}
  {\bibinfo  {journal} {Nano Letters}\ }\textbf {\bibinfo {volume} {17}},\
  \bibinfo {pages} {1496} (\bibinfo {year} {2017})}\BibitemShut {NoStop}%
\bibitem [{\citenamefont {Udvarhelyi}\ \emph {et~al.}(2018)\citenamefont
  {Udvarhelyi}, \citenamefont {Shkolnikov}, \citenamefont {Gali}, \citenamefont
  {Burkard},\ and\ \citenamefont {P\'alyi}}]{galiSpinStrain}%
  \BibitemOpen
  \bibfield  {author} {\bibinfo {author} {\bibfnamefont {P.}~\bibnamefont
  {Udvarhelyi}}, \bibinfo {author} {\bibfnamefont {V.~O.}\ \bibnamefont
  {Shkolnikov}}, \bibinfo {author} {\bibfnamefont {A.}~\bibnamefont {Gali}},
  \bibinfo {author} {\bibfnamefont {G.}~\bibnamefont {Burkard}},\ and\ \bibinfo
  {author} {\bibfnamefont {A.}~\bibnamefont {P\'alyi}},\ }\bibfield  {title}
  {\bibinfo {title} {Spin-strain interaction in nitrogen-vacancy centers in
  diamond},\ }\href {https://doi.org/10.1103/PhysRevB.98.075201} {\bibfield
  {journal} {\bibinfo  {journal} {Phys. Rev. B}\ }\textbf {\bibinfo {volume}
  {98}},\ \bibinfo {pages} {075201} (\bibinfo {year} {2018})}\BibitemShut
  {NoStop}%
\bibitem [{\citenamefont {Barfuss}\ \emph {et~al.}(2019)\citenamefont
  {Barfuss}, \citenamefont {Kasperczyk}, \citenamefont {K\"olbl},\ and\
  \citenamefont {Maletinsky}}]{maletinskyStrainTerms}%
  \BibitemOpen
  \bibfield  {author} {\bibinfo {author} {\bibfnamefont {A.}~\bibnamefont
  {Barfuss}}, \bibinfo {author} {\bibfnamefont {M.}~\bibnamefont {Kasperczyk}},
  \bibinfo {author} {\bibfnamefont {J.}~\bibnamefont {K\"olbl}},\ and\ \bibinfo
  {author} {\bibfnamefont {P.}~\bibnamefont {Maletinsky}},\ }\bibfield  {title}
  {\bibinfo {title} {Spin-stress and spin-strain coupling in diamond-based
  hybrid spin oscillator systems},\ }\href
  {https://doi.org/10.1103/PhysRevB.99.174102} {\bibfield  {journal} {\bibinfo
  {journal} {Phys. Rev. B}\ }\textbf {\bibinfo {volume} {99}},\ \bibinfo
  {pages} {174102} (\bibinfo {year} {2019})}\BibitemShut {NoStop}%
\bibitem [{\citenamefont {Glenn}\ \emph {et~al.}(2017)\citenamefont {Glenn},
  \citenamefont {Fu}, \citenamefont {Kehayias}, \citenamefont {{Le Sage}},
  \citenamefont {Lima}, \citenamefont {Weiss},\ and\ \citenamefont
  {Walsworth}}]{Glenn2017}%
  \BibitemOpen
  \bibfield  {author} {\bibinfo {author} {\bibfnamefont {D.~R.}\ \bibnamefont
  {Glenn}}, \bibinfo {author} {\bibfnamefont {R.~R.}\ \bibnamefont {Fu}},
  \bibinfo {author} {\bibfnamefont {P.}~\bibnamefont {Kehayias}}, \bibinfo
  {author} {\bibfnamefont {D.}~\bibnamefont {{Le Sage}}}, \bibinfo {author}
  {\bibfnamefont {E.~A.}\ \bibnamefont {Lima}}, \bibinfo {author}
  {\bibfnamefont {B.~P.}\ \bibnamefont {Weiss}},\ and\ \bibinfo {author}
  {\bibfnamefont {R.~L.}\ \bibnamefont {Walsworth}},\ }\bibfield  {title}
  {\bibinfo {title} {{Micrometer‐scale magnetic imaging of geological samples
  using a quantum diamond microscope}},\ }\href
  {https://doi.org/10.1002/2017GC006946} {\bibfield  {journal} {\bibinfo
  {journal} {Geochemistry, Geophysics, Geosystems}\ }\textbf {\bibinfo {volume}
  {18}},\ \bibinfo {pages} {3254} (\bibinfo {year} {2017})}\BibitemShut
  {NoStop}%
\bibitem [{\citenamefont {Kehayias}\ \emph {et~al.}(2019)\citenamefont
  {Kehayias}, \citenamefont {Turner}, \citenamefont {Trubko}, \citenamefont
  {Schloss}, \citenamefont {Hart}, \citenamefont {Wesson}, \citenamefont
  {Glenn},\ and\ \citenamefont {Walsworth}}]{StrainPaper}%
  \BibitemOpen
  \bibfield  {author} {\bibinfo {author} {\bibfnamefont {P.}~\bibnamefont
  {Kehayias}}, \bibinfo {author} {\bibfnamefont {M.~J.}\ \bibnamefont
  {Turner}}, \bibinfo {author} {\bibfnamefont {R.}~\bibnamefont {Trubko}},
  \bibinfo {author} {\bibfnamefont {J.~M.}\ \bibnamefont {Schloss}}, \bibinfo
  {author} {\bibfnamefont {C.~A.}\ \bibnamefont {Hart}}, \bibinfo {author}
  {\bibfnamefont {M.}~\bibnamefont {Wesson}}, \bibinfo {author} {\bibfnamefont
  {D.~R.}\ \bibnamefont {Glenn}},\ and\ \bibinfo {author} {\bibfnamefont
  {R.~L.}\ \bibnamefont {Walsworth}},\ }\bibfield  {title} {\bibinfo {title}
  {Imaging crystal stress in diamond using ensembles of nitrogen-vacancy
  centers},\ }\href {https://doi.org/10.1103/PhysRevB.100.174103} {\bibfield
  {journal} {\bibinfo  {journal} {Phys. Rev. B}\ }\textbf {\bibinfo {volume}
  {100}},\ \bibinfo {pages} {174103} (\bibinfo {year} {2019})}\BibitemShut
  {NoStop}%
\bibitem [{\citenamefont {Bar-Gill}\ \emph {et~al.}(2013)\citenamefont
  {Bar-Gill}, \citenamefont {Pham}, \citenamefont {Jarmola}, \citenamefont
  {Budker},\ and\ \citenamefont {Walsworth}}]{Bar-Gill2013}%
  \BibitemOpen
  \bibfield  {author} {\bibinfo {author} {\bibfnamefont {N.}~\bibnamefont
  {Bar-Gill}}, \bibinfo {author} {\bibfnamefont {L.~M.}\ \bibnamefont {Pham}},
  \bibinfo {author} {\bibfnamefont {A.}~\bibnamefont {Jarmola}}, \bibinfo
  {author} {\bibfnamefont {D.}~\bibnamefont {Budker}},\ and\ \bibinfo {author}
  {\bibfnamefont {R.~L.}\ \bibnamefont {Walsworth}},\ }\bibfield  {title}
  {\bibinfo {title} {{Solid-state electronic spin coherence time approaching
  one second.}},\ }\href {https://doi.org/10.1038/ncomms2771} {\bibfield
  {journal} {\bibinfo  {journal} {Nature communications}\ }\textbf {\bibinfo
  {volume} {4}},\ \bibinfo {pages} {1743} (\bibinfo {year} {2013})}\BibitemShut
  {NoStop}%
\bibitem [{\citenamefont {Bauch}\ \emph {et~al.}(2018)\citenamefont {Bauch},
  \citenamefont {Hart}, \citenamefont {Schloss}, \citenamefont {Turner},
  \citenamefont {Barry}, \citenamefont {Kehayias}, \citenamefont {Singh},\ and\
  \citenamefont {Walsworth}}]{P1DQ}%
  \BibitemOpen
  \bibfield  {author} {\bibinfo {author} {\bibfnamefont {E.}~\bibnamefont
  {Bauch}}, \bibinfo {author} {\bibfnamefont {C.~A.}\ \bibnamefont {Hart}},
  \bibinfo {author} {\bibfnamefont {J.~M.}\ \bibnamefont {Schloss}}, \bibinfo
  {author} {\bibfnamefont {M.~J.}\ \bibnamefont {Turner}}, \bibinfo {author}
  {\bibfnamefont {J.~F.}\ \bibnamefont {Barry}}, \bibinfo {author}
  {\bibfnamefont {P.}~\bibnamefont {Kehayias}}, \bibinfo {author}
  {\bibfnamefont {S.}~\bibnamefont {Singh}},\ and\ \bibinfo {author}
  {\bibfnamefont {R.~L.}\ \bibnamefont {Walsworth}},\ }\bibfield  {title}
  {\bibinfo {title} {Ultralong dephasing times in solid-state spin ensembles
  via quantum control},\ }\href {https://doi.org/10.1103/PhysRevX.8.031025}
  {\bibfield  {journal} {\bibinfo  {journal} {Phys. Rev. X}\ }\textbf {\bibinfo
  {volume} {8}},\ \bibinfo {pages} {031025} (\bibinfo {year}
  {2018})}\BibitemShut {NoStop}%
\end{thebibliography}%
\end{document}


\title{Supplemental Material for ``NV-Diamond Magnetic Microscopy using a Double Quantum 4-Ramsey Protocol''}

\date{\today}
\author{Connor A. Hart}  
\affiliation{\huPhys}
\affiliation{\MaryECE} \affiliation{\MaryPhys} \affiliation{\MaryQTC}

\author{Jennifer M. Schloss} 
\affiliation{\lincoln}
\affiliation{\mitPhys} 
\affiliation{\cbs}

\author{Matthew J. Turner} 
\affiliation{\huPhys}
\affiliation{\MaryECE} \affiliation{\MaryPhys} \affiliation{\MaryQTC}
\affiliation{\cbs}

\author{Patrick J. Scheidegger} 
\affiliation{\ethzurich}

\author{Erik Bauch} 
\affiliation{\cfa}

\author{Ronald L. Walsworth}  \affiliation{\huPhys}  \affiliation{\MaryECE}  \affiliation{\MaryPhys} \affiliation{\MaryQTC}
\affiliation{\cbs}
\affiliation{\cfa}

\renewcommand{\thetable}{S\arabic{table}}
\renewcommand{\thesection}{S\arabic{section}}   
\renewcommand{\theequation}{S\arabic{equation}}
\renewcommand{\thefigure}{S\arabic{figure}}
\renewcommand{\bibnumfmt}[1]{[S#1]}
\renewcommand{\citenumfont}[1]{S#1}
\renewcommand{\tablename}{\textbf{Supplemental Table}}
\renewcommand{\figurename}{\textbf{Supplemental Figure}}

\maketitle

\tableofcontents

\section{Crystal Stress Terms in the Ground State NV$^\text{-}$ Hamiltonian}

\noindent In this section, we describe the effect of crystal stress on the NV$^\text{-}$ ground state spin resonances. The ground state spin Hamiltonian for a chosen NV$^\text{-}$ orientation in the presence of crystal stress and a static (bias) magnetic field is given by~\cite{marcusStrainHam, galiSpinStrain, maletinskyStrainTerms}:
\begin{equation}\label{Eqn:suppNVham}
\begin{split}
H/h \approx & \left( D + M_{z} \right) S_{z}^2 + \frac{\gamma}{2\pi} \vec{B} \cdot \vec{S}_{} \\
& + M_{x} \left( S_{y}^2 - S_{x}^2 \right) \\
& + M_{y} \left( S_{x} S_{y} + S_{y} S_{x} \right) \\
& + N_{x} \left( S_{x} S_{z} + S_{z} S_{x} \right)  \\
& + N_{y} \left( S_{y} S_{z} + S_{z} S_{y} \right).  \\
\end{split} 
\end{equation} 
Here the z axis is defined along the NV$^\text{-}$ symmetry axis, $D \approx 2870$ MHz is the zero-field splitting, $S_{i}$ are the dimensionless spin-1 projection operators, $ \gamma/2\pi = g_{NV} \mu_B/h  \approx 28.03\,$GHz/T is the NV$^\text{-}$ gyromagnetic ratio, $\vec{B}$ is the magnetic field, and $M_{i}$  and $N_i$ are terms related to crystal stress. In the following analysis, we do not include the $N_i$ terms or contributions due to transverse magnetic fields, which are suppressed by the zero-field splitting~\cite{Glenn2017, StrainPaper, marcusStrainHam, galiSpinStrain, maletinskyStrainTerms}. 

The projection of the magnetic field along the NV$^\text{-}$ symmetry axis is indicated by $B_z$. We refer to $M_z$ as the axial crystal stress and define the transverse crystal stress, $M_\perp$, as  $-(M_x + i M_y)$. Using these definitions, Eq.~\ref{Eqn:suppNVham} can be rewritten, in matrix form, as:

\begin{equation}\label{eqn:suppNVHam2}
H/h = \left( \begin{array}{ccc}
D + M_z +\frac{\gamma}{2\pi} B_z & 0 & M_\perp \\
0 & 0 & 0 \\
M_\perp^* & 0 & D + M_z - \frac{\gamma}{2\pi} B_z
\end{array} \right).
\end{equation}

For non-zero $M_z$ and $M_\perp$, the energy eigenvalues of the NV$^\text{-}$ Hamiltonian (Eq.~\ref{eqn:suppNVHam2}) for the $|\pm 1\rangle$ states become

\begin{align}
E_{|\pm1\rangle}/h &= D + M_z \pm \sqrt{\left(\frac{\gamma}{2\pi} B_z\right)^2 +  \lvert\lvert M_{\perp} \rvert\rvert^2}\\ \label{eqn:suppNVener2}
&\approx  D + M_z \pm 
\left[\frac{\gamma}{2\pi} B_z 
+ \frac{\lvert\lvert M_{\perp} \rvert\rvert^2}{2 \frac{\gamma}{2\pi} B_z}
+ \mathcal{O}\left(\frac{\lvert\lvert M_{\perp} \rvert\rvert^4}{B_z^2} \right)
\right].
\end{align}

To first order, crystal stress shifts the energies of the $\ket{\pm1}$ states in common mode via the axial-stress term $M_{z}$. Single quantum (SQ) measurements, which probe the difference between $E_{\ket{0}}$ and either $E_{\ket{\pm1}}$, are vulnerable to these axial-stress-induced shifts. Conversely, double quantum (DQ) measurements, which probe the difference between $E_{\ket{+1}}$ and $E_{\ket{-1}}$, are insensitive to axial stress. 

At second order, the transverse crystal stress term $M_\perp$ induces magnetic-like, differential shifts in the energy eigenvalues $E_{\ket{\pm1}}$, which are not mitigated in either the SQ or DQ sensing bases. However, these transverse-stress-induced shifts are suppressed by a factor $\vert \vert M_\perp \vert \vert / \left(\gamma B_z/\pi \right)$ where, as noted earlier, $B_z$ is the projection of $\vec{B}$ along a particular NV$^\text{-}$ orientation. In the present work, if we assume variations in $M_z\!\approx\!M_\perp\! \approx\,$\SI{100}{\kilo\hertz}, then the aligned bias magnetic field of \SI{5}{\milli\tesla} in the present experiment provides a suppression of $\approx\,$450. The resultant shift of \SI{0.2}{\kilo\hertz} (for $M_\perp\!=\,$\SI{100}{\kilo\hertz}) is approximately equivalent to a magnetic field of \SI{8}{\nano\tesla} -- sufficiently small to be negligible compared to the axial-stress-induced shifts that dominate the SQ measurements presented in the main text. As apparent from the above analysis, the effects of transverse stress inhomogeneity can be further suppressed by increasing $B_z$. 

\section{Sources of Microwave Pulse Errors}

\textit{Spatial Rabi Gradients} -- Figure~1(c) in the main text depicts the Rabi gradient for a mm-scale shorted coaxial loop. Although the spatial properties of the applied MW field depend upon setup-specific synthesis and delivery approaches, the $\pm4$\% ($\pm200\,$kHz) gradient on top of an average Rabi frequency $\Omega_\text{avg}\!=\!5\,$MHz shown in the main text is typical. 

\textit{Hyperfine Splitting-Induced Detunings} -- The NV$^\text{-}$ hyperfine splitting can also introduce MW pulse errors as a result of the detuning-dependent effective Rabi frequency $\Omega_\text{eff}$. Samples grown with the naturally abundant, $^{14}\text{N}$ nitrogen isotope ($I\!=\!1$) are particularly vulnerable because any single-tone MW field is detuned from at least two of the three hyperfine-split resonances. The effective Rabi frequency is given by $\Omega_\text{eff}\!=\!\sqrt{\Omega^2 + \Delta^2}$, where $\Omega$ is the on-resonance Rabi frequency and $\Delta$ is the detuning from a particular hyperfine resonance. In this work, the MW pulse errors induced by the hyperfine splitting ($\Omega_\text{avg}\!=\!5\,$MHz, $\Delta\!\approx\!2.2\,$MHz) result in effective Rabi frequencies of $\Omega_\text{eff}\!=\!5\pm0.46\,$MHz for the two detuned hyperfine resonances ($m_I\!=\!\pm1$), approximately double the Rabi gradient of $\pm200\,$kHz. Potential approaches to mitigate hyperfine-induced pulse errors include increasing the MW Rabi frequency, implementing pulse envelope shaping, or operating equally detuned from the two hyperfine split resonances of isotopically-purified $^{15}\text{N}$ ($I\!=\!1/2$). All three of these directions present challenges, adding experimental complexity or requiring changes to the CVD synthesis process.

\textit{Stress-Induced Detunings} -- Finally, axial-stress-induced shifts in the NV$^\text{-}$ resonance frequencies can also result in effective Rabi frequency variations. Using the values reported in Sec. IV as an estimate ($|D_{90}-D_{10}|\!=\,$280\,(14)\,kHz), the effective Rabi frequency deviates from $\Omega_\text{avg}\!=\!5\,$MHz by only about 10$\,$kHz. However, in pixels with larger stress-induced shifts, the effect would become comparable to the two other sources described above. 

\section{Generalized 4-Ramsey Phase Requirements}

\begin{figure}[htbp]
 \begin{center}
 \begin{overpic}[width=1\columnwidth]{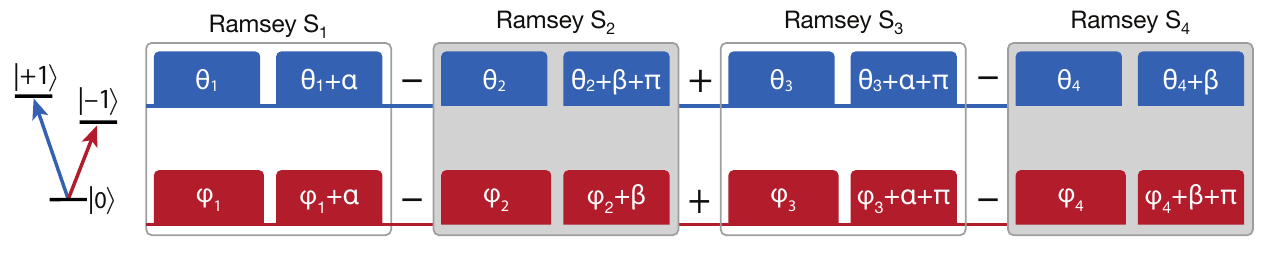}
 \end{overpic}
 \end{center}
 \caption{\label{4r_phases}
 Generalized phase requirements for the DQ 4-Ramsey protocol. The phases for $\alpha$, $\beta$, $\theta_i$, and $\phi_i$ for $i=1-4$ may be chosen arbitrarily. $S_1$ and $S_3$ yield DQ signals which differ in sign from $S_2$ and $S_4$ when sensing the same magnetic field.}
 \end{figure}

\noindent We describe here the generalized phase requirements for the DQ 4-Ramsey protocol (refer to Fig.~\ref{4r_phases}). First, only the relative phase shifts between the two MW tones addressing the $\ket{0}\rightarrow \ket{-1}$ and $\ket{0}\rightarrow\ket{+1}$ transitions are critical to the DQ 4-Ramsey protocol. As a consequence, the phases $\theta_i$ and $\phi_i$ in Fig.~\ref{4r_phases} can be selected arbitrarily for the initial two-tone MW pulse in the each sequence $S_i$. The choices of $\alpha$ and $\beta$ for the final MW pulses in each sequence can also be chosen without restriction (in the main text: $\alpha=\beta=0$). However, subsequent MW pulses must have a specific phase relationship across $S_{1-4}$ as depicted in Fig~\ref{4r_phases} by the $\pi$ phase shifts for select pulses. Second, the order of $S_{1-4}$ can be permuted. For example, $S_1$ and $S_3$ may be swapped without altering $S_{4R}
= S_1 - S_2 + S_3 - S_4$. In this work, the form of $S_{4R}$ was chosen to be compatible with the pixel-by-pixel, lock-in detection scheme of the heliCam C3 camera (see Supplemental Sec.~S5). 
 
\section{SQ and DQ 2-Ramsey Protocols}

\noindent This section provides additional details on the conventional SQ and DQ 2-Ramsey phase alternation protocols. The SQ 2-Ramsey single-channel (photodiode) data presented in Fig.~3(b) and 3(d) in the main text is calculated using the visibility $S_{2R} = (S_1 - S_2) / (S_1+S_2)$ where $S_i$ indicates the signal produced by measurement $i = 1,2$. As defined above, $S_{2R}$ suppresses noise sources that vary slowly compared to the time over which $S_1$ and $S_2$ are acquired~\cite{Bar-Gill2013}. In contrast, when the SQ 2-Ramsey protocol is implemented using the heliCam C3 camera, only the numerator of $S_{2R}$ is accessible (see further discussion in Supplemental Sec.~S5).

As with the SQ results, the single-channel (photodiode) DQ 2-Ramsey data presented in Fig.~3(b) and 3(d) in the main text is also calculated according to the visibility $S_{2R} = (S_1 - S_2) / (S_1+S_2)$. The MW pulse phases chosen for the DQ 2-Ramsey measurements are identical to those of the first two Ramsey signals ($S_1$ and $S_2$) in the 4-Ramsey protocol. 
 
Comparing the DQ 2-Ramsey and 4-Ramsey representations in Fig.~2(b,c) of the main text, it is apparent that calculating $S_{2R}$ requires half the number of Ramsey signals as $S_{4R}$. For implementations where each Ramsey measurement is accessible, the bandwidth of the 2-Ramsey protocols is thus a factor of two greater than that of the 4-Ramsey protocol. However, because the DQ 4-Ramsey protocol acquires Ramsey signals at the same rate as the DQ 2-Ramsey protocol (which all add constructively), there is no corresponding impact on the magnetic sensitivity. Additionally, if desired, the 4-Ramsey data could be post-processed as 2-Ramsey data using $S_{2R}$ to recover the sacrificed bandwidth.

 
\section{Ramsey Imaging Using the heliCam C3 Lock-in Camera}
 
\noindent The heliCam C3 operates by subtracting alternating exposures in analog detection; and then digitizing the background-subtracted signal, enabling the detected magnetic field information to fill each pixel's 10-bit dynamic range. Each frame from the camera contains the accumulated difference signal for multiple exposure pairs. The device is capable of operating at internal demodulation frequencies of up to 250$\,$kHz (exposure rates of up to 1$\,$MHz). 

The demodulation frequency can be defined by a user-input demodulation signal (square wave) with each demodulation cycle broken into four quarters. The exposures from the first and third quarters subtract to produce an in-phase image on the camera's I channel. Similarly, the exposures from the second and fourth quarters subtract to produce a quadrature image on the camera's Q channel. The output from the camera consists of these two images per frame. Since each demodulation cycle includes four exposures, the internal exposure rate is $4\times F_\text{demod}$ where $F_\text{demod}$ indicates the inverse of one demodulation cycle period.

\begin{figure}[htp]
 \begin{center}
 \begin{overpic}[width=1\columnwidth]{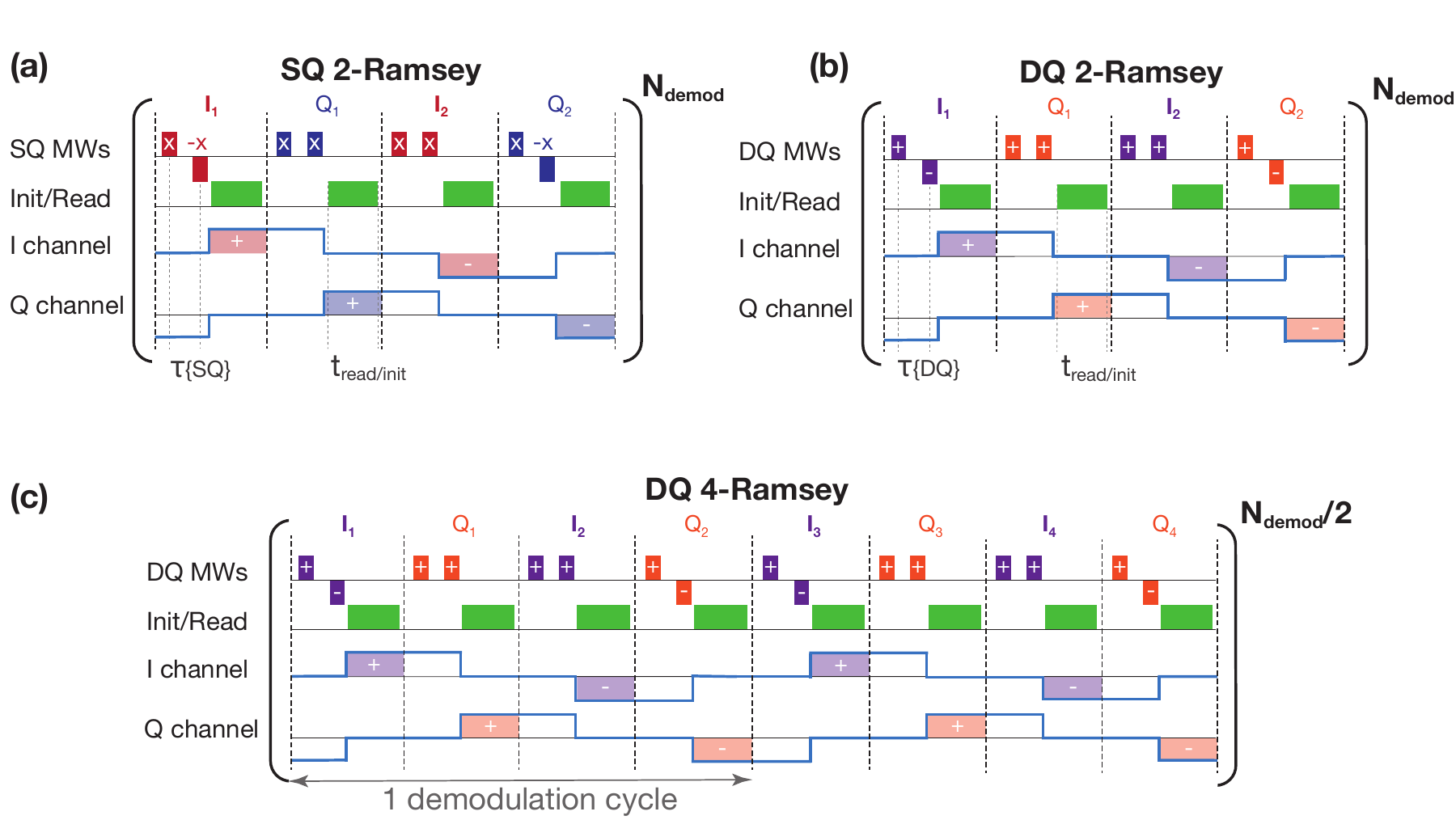}
 \end{overpic}
 \end{center}
 \caption{\label{helicam}
 \textbf{Schematics of measurement protocols used for magnetic imaging.}  
\textbf{(a)} A single quantum (SQ) 2-Ramsey protocol is shown synchronized with detection by the heliCam C3 lock-in camera. The single demodulation cycle depicted is repeated N$_\text{demod}$ times within each frame. Two SQ 2-Ramsey (red and blue pulses) protocols are interwoven to generate images on the I and Q channels. The camera exposures (shaded regions on the I and Q channels) are chosen to collect the fluorescence from the optical readout/initialization pulses (green pulses). The optical pulses are contained within the exposure duration.
For optimal magnetic sensitivity, the Ramsey sensing duration, $\tau\{\text{SQ}\} = 910\,$ns, is approximately the median $T_2^*\{\text{SQ}\}$ across the field of view. The duration of the combined readout and initialization optical pulses ($t_\text{init/read}=4\,\upmu$s) is experimentally determined to yield the optimal magnetic sensitivity. \textbf{(b)} Schematic of a DQ 2-Ramsey protocol. The optimal sensing duration is approximately given by the median $T_2^*\{\text{DQ}\}$ ($\tau\{\text{DQ}\} = 662\,$ns). \textbf{(c)} Schematic of a double quantum (DQ) 4-Ramsey protocol implemented on the helicam. The two DQ 4-Ramsey protocols (orange and purple pulses), which are interwoven to generate images on the I and Q channels, span two demodulation cycles. However, the total number of demodulation cycles per frame remains constant compared to the 2-Ramsey protocols.}
 \end{figure}
 
The I and Q images produced during each frame contain the accumulated difference signal from $N_\text{demod}$ repetitions of the demodulation cycle described above. In the present work, the typical number of demodulation cycles ($N_\text{demod}=24$) and demodulation rates ($F_\text{demod} \approx 35\,$kHz) result in external frame rates ($F_\text{S}$) of approximately 1.5$\,$kHz. The maximum external frame rate of the camera is 3.8$\,$kHz. Note that in the main text, the exposure rate of ($4\times F_\text{demod} \approx 140\,$kHz) is referenced instead of $F_\text{demod}$.


For the magnetic sensitivity discussion in Sec.~V, the SQ and DQ external frame rates differ slightly ($F_S\{\text{SQ}\} = 1.45\,$kHz versus $F_S\{\text{DQ}\} = 1.5\,$kHz) because the optimal sensing duration is approximately given by the median $T_2^*$ and therefore depends upon the sensing basis ($\tau\{\text{SQ}\} = 910\,$ns and $\tau\{\text{DQ}\} = 662\,$ns). Additionally, accounting for the four optical pulses ($t_\text{init/read}=4\,\upmu$s each), four pairs of microwave pulses ($<1\,\upmu$s per cycle), and delays between control pulses (8$\,\upmu$s per cycle), the duration of each demodulation cycle, $t_\text{demod}$, is 28.7$\,\upmu$s (27.8$\,\upmu$s) for the SQ (DQ) measurements.
 
Figure~\ref{helicam} depicts the implementation of pulsed NV$^\text{-}$ readout using the heliCam in further detail. The MW and optical control pulses are synchronized with the camera's I and Q exposures. Two protocols are interleaved such that both the I and Q channels produce images with the proper phase alternation (e.g. SQ 2-Ramsey, DQ 2-Ramsey, DQ 4-Ramsey). 
Additionally, the MW pulse phases for the Q channel are chosen such that the Q channel contains the negative magnetic signal compared to the I channel image. $S_I$ is then subtracted from $S_Q$ to yield a single image per frame.

While a full 2-Ramsey protocol requires only one demodulation cycle, compared to two cycles for the 4-Ramsey protocol, the total number of demodulation cycles per frame is fixed to $N_\text{demod} = 24$ for both the 2-Ramsey and 4-Ramsey protocols. As a consequence, the SQ and DQ frame rates are the same (neglecting the few percent change due to the basis-dependent sensing interval). Additionally, the difference signals between exposure pairs add constructively for the duration of the frame. Thus, the choice of protocol does not impact bandwidth or sensitivity when using the heliCam C3.


\section{Ramsey Fringe Imaging}

\begin{figure}[htbp]
 \begin{center}
 \begin{overpic}[width=1\columnwidth]{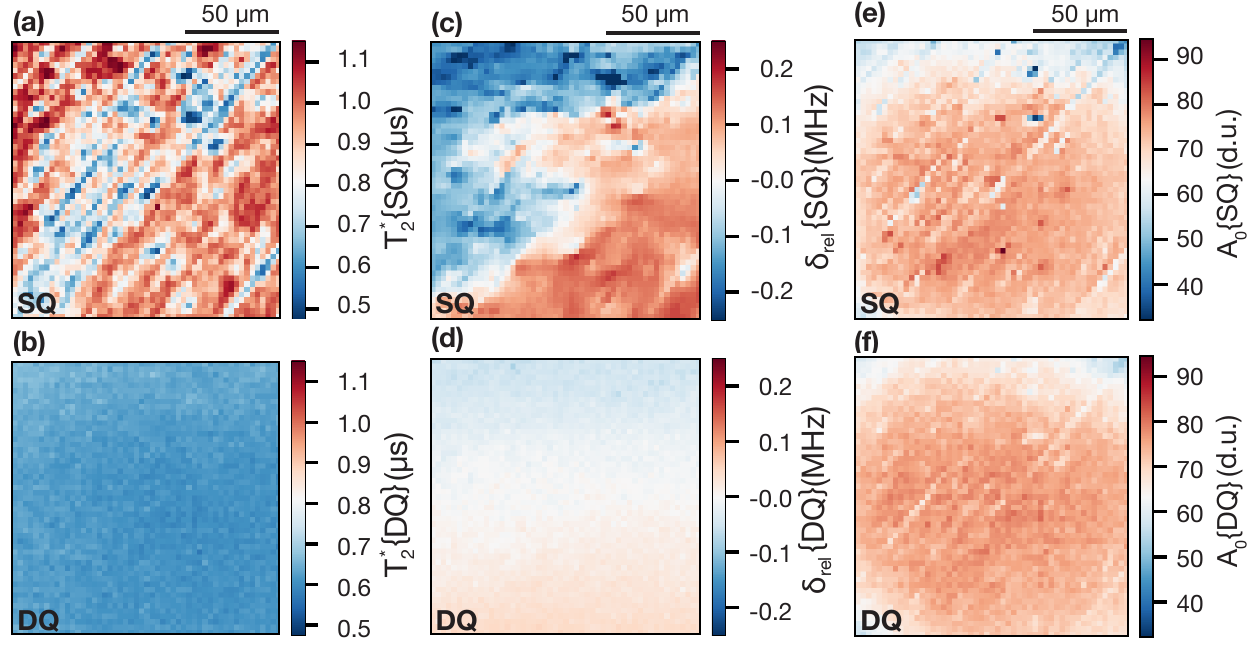}
 \end{overpic}
 \end{center}
 \caption{\label{fringe_all}
 \textbf{Imaging NV$^\text{-}$ Ensemble Spin Properties. (a)} Image of the single quantum (SQ) $T_2^*$ extracted by fitting the Ramsey free induction decay to Eq.~4 in the main text. The field of view is \SI{125}{\micro\meter} by \SI{125}{\micro\meter}. Spatial variations in $T_2^*$\{\text{SQ}\} are attributed to stress-induced broadening of the NV$^\text{-}$ resonances within the 3-dimensional volume imaged onto a pixel. 
 \textbf{(b)} Image of the double quantum $T_2^*$\{DQ\} measured using the 4-Ramsey protocol across the same field of view as shown in (a). In pixels, with minimal stress gradients, the $T_2^*$\{DQ\} is half the $T_2^*$\{SQ\}, as expected, due to the effectively doubled dipolar coupling to the surrounding paramagnetic spin bath~\cite{P1DQ}, which dominates NV$^\text{-}$ dephasing.
 \textbf{(c)} Image of the relative SQ resonance shifts $\delta_\text{rel}$\{SQ\} away from the median SQ Ramsey fringe frequency $f_\text{SQ}$. Variations in $\delta_\text{rel}$\{SQ\} are attributed to axial-stress-induced shifts of the NV$^\text{-}$ resonance frequencies between pixels.
 \textbf{(d)} Image of the relative DQ detuning $\delta_\text{rel}$\{DQ\}. The axial-strain-induced shifts apparent in (c) are mitigated. Inhomogeneity in the applied bias magnetic field $B_0$ results in a residual gradient of less than 1.4$\,\upmu$T (\SI{40}{\kilo\hertz} after accounting for the doubled gyromagnetic ratio in the DQ sensing basis).
 \textbf{(e)} Image of the SQ Ramsey fringe amplitude $A_0$\{SQ\} corresponding to the spin transition associated with the $m_I\!=\!0$ nuclear spin state, in digital units (D.U.) as reported by the heliCam. 
 \textbf{(f)} Image of the DQ Ramsey fringe amplitude $A_0$\{DQ\} corresponding to the spin transition associated with the $m_I = 0$ nuclear spin state, in digital units (D.U.) as reported by the heliCam.
 }
 \end{figure}

\noindent In Figs.~\ref{fringe_all}(a)-\ref{fringe_all}(d), the SQ and DQ dephasing times $T_2^*$ and relative resonance shifts $\delta_\text{rel}$ are reproduced from Fig.~3 in the main text. In addition, the extracted $A_0$ fringe amplitudes for the SQ and DQ sensing bases are shown in Figs.~\ref{fringe_all}(e) and \ref{fringe_all}(f). As noted in the main text, the extracted amplitudes $A_{i}$ (indexed by $m_I ={-1,0,1}$) for the measured Ramsey fringes are proportional to the measurement contrast and reported in digital units (D.U.). The comparable median amplitudes of $A_0\{\text{SQ}\}$ and $A_0\{\text{DQ}\}$ ($72.1$ and $73.5$ in digital units as reported by the camera) and relative inter-decile ranges ($21\,$\% and $14\,$\%) are largely due to the Gaussian intensity profile of the excitation beam and fixed exposure duration. Shorter length-scale spatial variations ($<\,$\SI{50}{\micro\meter}) in the extracted SQ amplitude exhibit a strong correlation with images of the $T_2^*$\{SQ\} and relative resonance shifts $\delta_\text{rel}$\{SQ\} due to stress gradients decreasing the ODMR contrast and inducing variations in the effective Rabi frequency.

\section{Per-Pixel Allan Deviation}
 
\begin{figure}[htbp]
 \begin{center}
 \begin{overpic}[width=0.6\columnwidth]{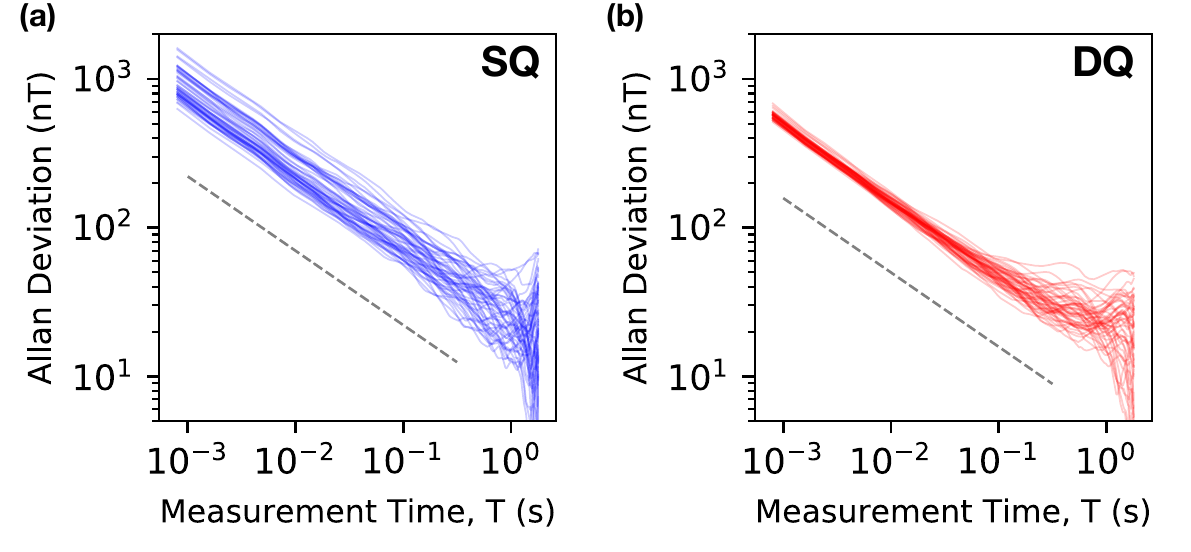}
 \end{overpic}
 \end{center}
 \caption{\label{allandev} Allan deviations for magnetic field measurements as a function of measurement time. (a) Measurements using a SQ 2-Ramsey sensing protocol for 50 randomly selected pixels across the same field of view as used for measurements shown in Fig.~4 of the main text. (b) Measurements using the DQ 4-Ramsey sensing protocol for the same 50 pixels as depicted in (a). The free precession interval and detuning are optimized for each sensing basis to minimize the median per-pixel magnetic sensitivity. Dashed grey lines depict a power law scaling proportional to $T^{-1/2}$ as a guide to the eye.}
 \end{figure}

\noindent For the magnetic sensitivities reported in the main text and depicted in Fig.~4, SQ 2-Ramsey and DQ 4-Ramsey data was collected, each for 1 second of measurement time and neglecting the overhead time required to transfer data from the camera to the computer. Allan deviations calculated using similar SQ 2-Ramsey and DQ 4-Ramsey measurements (with greater than \SI{1}{\second} of data) are shown in Fig.~\ref{allandev} for 50 randomly selected pixels across the same field of view presented in the main text. For both measurement protocols, the Allan deviation exhibits a power law scaling proportional to $T^{-1/2}$, where $T$ is the measurement time. The greater inhomogeneity in SQ magnetic sensitivities is apparent from the larger spread in SQ measurements compared to DQ measurements. 

\section{Averaged Magnetic Images}

\begin{figure}[htbp]
 \begin{center}
 \begin{overpic}[width=0.8\columnwidth]{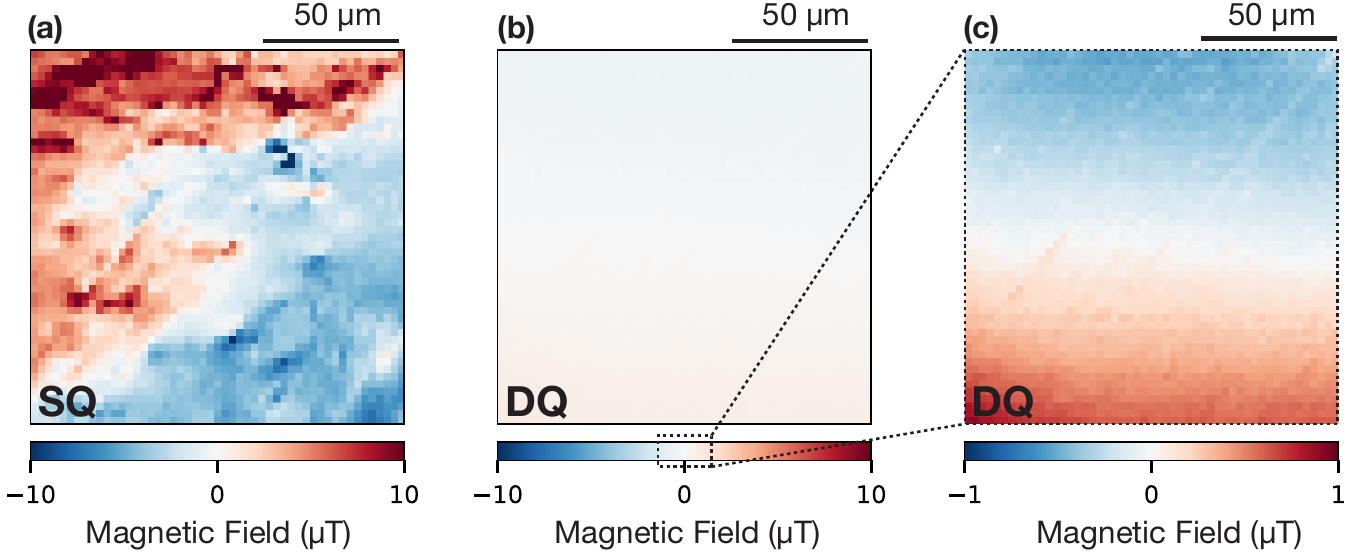}
 \end{overpic}
 \end{center}
 \caption{\label{b0flat}
    Averaged SQ 2-Ramsey and DQ 4-Ramsey images when operating at optimal sensing conditions across the same field of view as shown in Fig.~\ref{fringe_all} and \ref{allandev}. (a) SQ image exhibiting \SI{10}{\micro\tesla}-scale spatial variations due to stress-induced NV$^\text{-}$ resonance shifts. (b) DQ image with reduced spatial variations due to insensitivity to non-magnetic sources. (c) Same data as (b) but with a $10\times$ reduced magnetic field scale to highlight residual spatial variation predominantly correlated with the bias magnetic field, which is aligned along one NV$^\text{-}$ axis at an angle of 54.7$^\circ$ relative to the normal to the image plane.}
 \end{figure}

\noindent Figure~\ref{b0flat} compares the magnetic field images produced after acquiring 1 second of data (each) using the SQ 2-Ramsey protocol and DQ 4-Ramsey protocols. No additional magnetic sources are applied beyond the \SI{5}{\milli\tesla} bias magnetic field. However, the SQ image in Fig.~\ref{b0flat}(a) exhibits large non-magnetic spatial variations on the order of \SI{10}{\micro\tesla} due to axial-stress inhomogeneity. The inability of SQ measurements to distinguish axial crystal stress from magnetic fields complicates the analysis of such images and necessitates large measurement dynamic range (at the cost of magnetic sensitivity). In contrast, the pixels in the DQ image [Fig.~\ref{b0flat}(b)] respond only to magnetic-sources (to leading order), such that the resulting DQ image exhibits an order of magnitude reduced variation [as highlighted by the reduced color-scale in Fig.~\ref{b0flat}(c)]. 

The residual variation in the DQ magnetic image [Fig.~\ref{b0flat}(c)] is predominantly correlated with inhomogeneity in the bias magnetic field, which can be further engineered to reduce residual magnetic gradients. However, subtle scratch-like features on the sub-100$\,$nT scale are also visible, stretching diagonally across the field of view. These features are attributed to damage on the surface of the diamond substrate before growth of the nitrogen-doped layer. Similar features can be identified in the extracted Ramsey fringe amplitudes in Figs.~\ref{fringe_all}(c) and \ref{fringe_all}(d) for the SQ and, to a lesser extent, DQ sensing bases. The persistence of these features in both bases is informative and suggests that transverse-stress gradients, which are not suppressed by either sensing basis, may contribute to the residual variation of the averaged DQ image shown in Figs.~\ref{b0flat}(b) and \ref{b0flat}(c) (see Supplemental Sec.~1).

\clearpage
\bibliography{DQ4RSupplemental}